\documentclass[a4paper,11pt]{article}
\pdfoutput=1 

\usepackage{jheppub} 

\usepackage[T1]{fontenc} 
\usepackage{caption}
\usepackage{subcaption}
\usepackage{ptdr-definitions}
\usepackage[colorinlistoftodos]{todonotes}

\title{Explainable AI for ML jet taggers using expert variables and layerwise relevance propagation}

\author[a]{Garvita Agarwal,}
\author[a]{Lauren Hay,}
\author[a]{Ia Iashvili,}
\author[a,b]{Benjamin Mannix,}
\author[a]{Christine McLean,}
\author[a]{Margaret Morris,}
\author[a]{Salvatore Rappoccio,}
\author[a,c]{Ulrich Schubert}

\affiliation[a]{Physics Department, and Institute for Computational and Data Sciences, University at Buffalo, State University of New York, Buffalo, New York, USA}
\affiliation[b]{Department of Physics, University of Oregon, Eugene, OR, USA}
\affiliation[c]{Google, Pittsburg, PA, USA}

\emailAdd{garvitaa@buffalo.edu}
\emailAdd{lmhay@buffalo.edu}
\emailAdd{iashvili@buffalo.edu}
\emailAdd{brmannix@buffalo.edu}
\emailAdd{ch.mclean@cern.ch}
\emailAdd{morris35@buffalo.edu}
\emailAdd{srrappoc@buffalo.edu}
\emailAdd{ulrichsc@buffalo.edu}

\abstract{A framework is presented to extract and understand decision-making information from a deep neural network (DNN) classifier of jet substructure tagging techniques. The general method studied is to provide expert variables that augment inputs (“eXpert AUGmented” variables, or XAUG variables), then apply layerwise relevance propagation (LRP) to networks both with and without XAUG variables. The XAUG variables are concatenated with the intermediate layers after network-specific operations (such as convolution or recurrence), and used in the final layers of the network. The results of comparing networks with and without the addition of XAUG variables show that XAUG variables can be used to interpret classifier behavior, increase discrimination ability when combined with low-level features, and in some cases capture the behavior of the classifier completely. The LRP technique can be used to find relevant information the network is using, and when combined with the XAUG variables, can be used to rank features, allowing one to find a reduced set of features that capture part of the network performance. In the studies presented, adding XAUG variables to low-level DNNs increased the efficiency of classifiers by as much as 30-40\%. In addition to performance improvements, an approach to quantify numerical uncertainties in the training of these DNNs is presented. 
}

\begin{document} 
\maketitle
\flushbottom

\section{Introduction}
\label{sec:intro}

Machine learning (ML) has become an extremely prevalent tool in the classification of hadronically decaying highly Lorentz-boosted objects ("boosted jets") and to study the internal structure of hadronic jets ("jet substructure")~\cite{Abdesselam:2010pt,Altheimer:2012mn,Altheimer:2013yza,Adams:2015hiv,Larkoski:2017jix,Asquith:2018igt,Kasieczka:2019dbj}. In these areas, classification tasks are common, for which artificial neural networks (ANNs) are well suited. Recent work in deep neural networks (DNNs) has shown tremendous improvements in identification of boosted jets over selections based on expert variables (see Ref.~\cite{Larkoski:2017jix} and references therein). These algorithms typically make use of classifiers based on convolutional neural networks (CNNs) and recurrent neural networks (RNNs). 

However, these improvements come at a significant cost. The underlying understanding of particular decisions is lost (although it could be argued that this is not a drawback). This paper provides a  method to elucidate classifier decisions in an explainable framework. This can assist in understanding of systematic uncertainties, as well as to develop better expert variables to capture the behavior of the classifier in a simpler way. We will also demonstrate that combining expert features with low-level information can improve network performance, similar (but not identical) to the approach in Ref.~\cite{Guest:2016iqz}. 

Collectively, explaining classifiers in artificial intelligence (AI) is referred to as "eXplainable AI" (XAI); see Refs.~\cite{Murdoch_2019,DBLP:journals/corr/Lipton16a} for reviews. 
The classifiers we will investigate are CNNs and RNNs~\cite{Goodfellow-et-al-2016}. CNNs utilize a (usually) fixed-size input vector, processed by convolutions with various filters to highlight features such as polarity, divergence, curl, etc. These higher-level processing features are then combined with a dense output layer, where the final classification is performed. In this paper, we will only consider binary classification ("signal" versus "background"), although this is not a restriction. RNNs are ordinarily not restricted to fixed-size inputs, allowing for arbitrary lengths of inputs. The key element is that the output of some or all output nodes is then connected back into the inputs, allowing for a temporal state machine to be naturally implemented. The inputs to RNNs are not restricted and can be of any length. Traditional multivariate techniques in HEP are restricted to use so-called "expert" variables, which are high-level features of a dataset that are able to classify between events. More recent developments utilize lower-level features, such as measurements of particle momenta and energy, or measurement features from detectors. Our approach will be to combine these features in an attempt to explain classifiers based on lower-level information.

There are many options to explain individual classifier decisions, mostly based on local approximations of the classifier function. Examples include Local Interpretable Model-Agnostic Explanations (LIME)~\cite{DBLP:journals/corr/RibeiroSG16}, Sparse LInear Subset Explanations (SLISE)~\cite{SLISE}, Layerwise Relevance Propagation (LRP)~\cite{10.1371/journal.pone.0130140,DBLP:journals/corr/abs-1708-08296,Montavon2019}, and explanation vectors~\cite{cf736d955d6e4296a9b7255bfee3b403}. We will utilize the LRP method in this paper due to simplicity of use and interpretation, although others could be used as well. 

\begin{figure}[!ht]
    \centering
    \includegraphics[width = \textwidth]{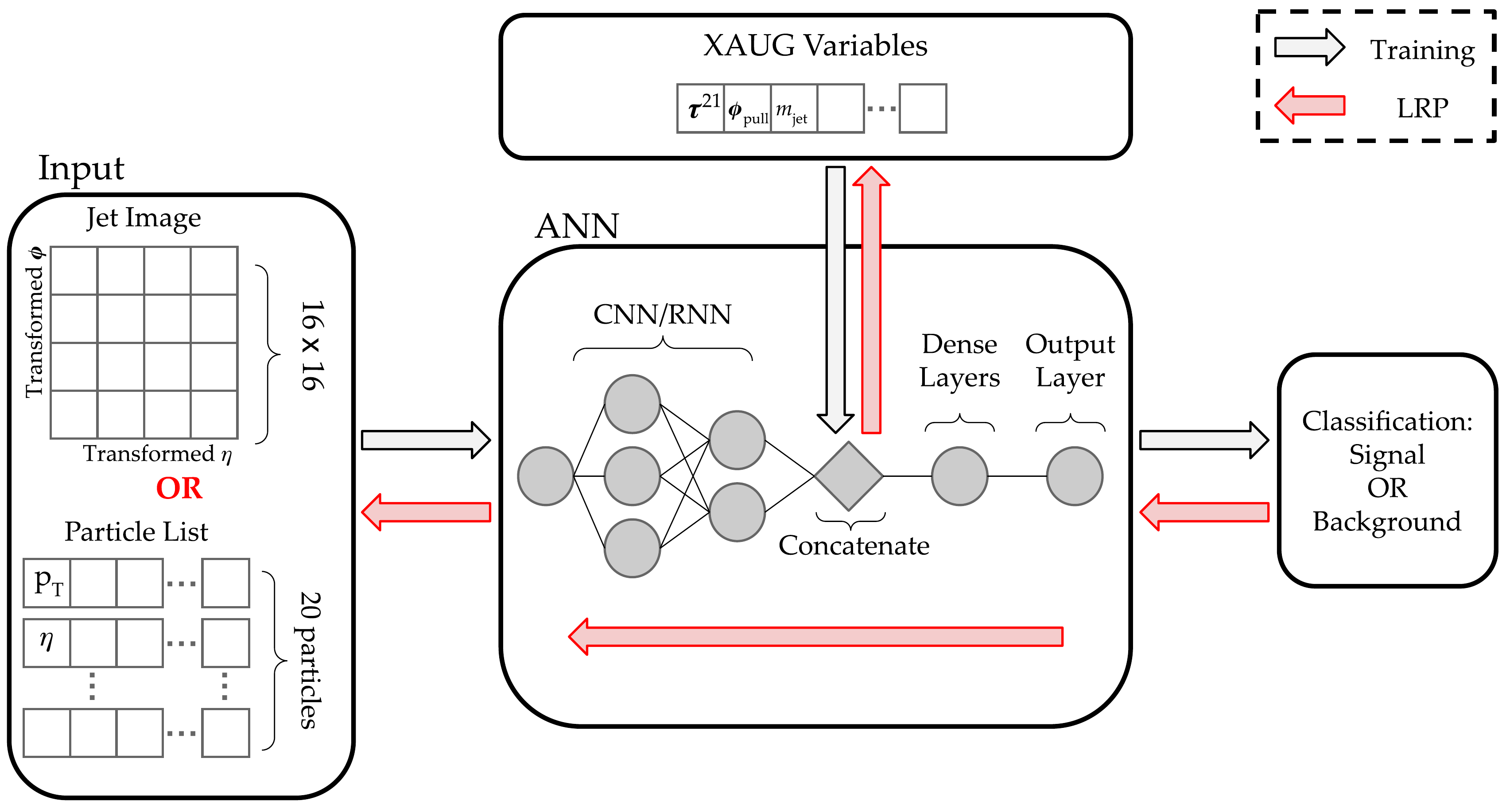}
    \caption{A schematic representation of our XAUG variables with LRP method.}
    \label{fig:block-diag}
\end{figure}

Particle physics has a unique property, in that many of the observables we are interested in have a full theoretical framework that can predict, or at least describe, their distributions. In particular, the expert variables provided can often be shown to fully capture available kinematic information, as is done in Refs.~\cite{Datta:2017rhs,Lim:2018toa,Chakraborty:2019imr,Chen:2019apv}, or even have a theoretical description of the classifier itself~\cite{Kasieczka:2020nyd}. However, there are other observables for which the theoretical completion is lower than others. In particular, information that can determine the original progenitor of the jet (such as flavor information) is sometimes less theoretically developed. Many new identifiers that show extremely high levels of performance rely on this type of information, such as track impact parameters, secondary vertex information, lepton content, and particle multiplicities, for instance as shown in Ref.~\cite{Aad:2019aic,Aad:2015ydr,Aaboud:2018xwy,Aad:2019uoz,Sirunyan:2017ezt,CMS:2019gpd}. Some of these observables are not easily calculable, and many are not even infrared/collinear safe (see, for instance, Ref.~\cite{Salam:2009jx} for a discussion of infrared and collinear safety). As such, further tools are currently needed to extricate the various sources of discrimination aside from theoretical calculations.

Our approach is to augment low-level input information with high-level expert variables, referred to as "eXpert-AUGmented" (XAUG) variables, and utilize a local approximation system such as LRP to understand the importance of various pieces of information as shown in Figure \ref{fig:block-diag}. The XAUG variable approach is similar to other contexts such as Ref.~\cite{Guest:2016iqz,Kasieczka:2017nvn,Moreno:2019neq,Moreno:2019bmu,Mikuni:2020wpr}, 
however, our strategies and goals are different. 
Instead of using expert variables alongside low-level features in the inputs, we augment the entire low-level network with expert features by concatenating them to the post-processed decisions as a whole (either after the convolution or recurrent network decisions). This allows a simple network to investigate and learn high-level (i.e. more abstract) details in the inputs.  
We also seek to provide these XAUG variables for dimensionality reduction in the function space of the DNN classifier, similar in strategy to the analytic tools proposed in~\cite{Datta:2017rhs}, but with an eye toward extrication of the experimentally-available information that may not be easily captured by theory, such as flavor information. This can assist in explaining individual classifier decisions. In an ideal case, the DNN classifier can in fact be fully contained within the XAUG variables in a simple multilayer perceptron (MLP).

We develop a framework that can be applied to any classifier in an attempt to reduce the dimensionality of the problem by 
replacing or augmenting the lower-level/higher-dimensional
features (i.e. many inputs per jet) by higher-level/lower-dimensional XAUG variables (i.e. one input per jet).
We will show that the behavior of the DNN can be captured robustly by the addition of appropriate XAUG variables, and in some cases can entirely capture the behavior. This technique is not limited to information that can be theoretically described or predicted. 

Firstly, we develop a trivial "toy" model with only a few features that are fully captured by XAUG variables. We show that the classifier decisions of such a DNN will be the same as a simpler classifier based only on XAUG variables themselves. We developed a 2D CNN based on "images" similar (but not identical) to the approaches in Refs.~\cite{Cogan:2014oua,deOliveira:2015xxd,Kasieczka:2017nvn}, as well as a 1D CNN and a 1D RNN based on particle lists inspired by the algorithm inputs in Ref.~\cite{Guest:2016iqz,CMS:2019gpd}.

Secondly, we develop several classifiers to distinguish boosted $Z$ bosons decaying to closely separated $b\bar{b}$ pairs ($Z\rightarrow b\bar{b}$) from standard QCD jets based on simulations. Once again, we investigate several cases, including a 2D jet image-based CNN (using only jet kinematic and shape information), a 1D particle-list CNN, and a 1D particle-list RNN. The latter two contain information beyond the kinematics of the jet, such as particle content and decay impact parameters. 

As mentioned above, it has been shown in 
Refs.~\cite{Datta:2017rhs,Lim:2018toa,Chakraborty:2019imr,Chen:2019apv} that the kinematic information of such classifiers is exhausted by angularity variables. One example "basis" is the set of $N$-subjettiness variables~\cite{Thaler:2010tr,Thaler:2011gf}. As such, classifier decisions with only kinematic information should be almost entirely correlated with existing kinematic expert variables when those variables are used as XAUG variables. Other information is not captured by these kinematic observables, however, such as the flavor and soft radiation in the jet such as the "jet pull"~\cite{Gallicchio:2010sw}. An advantage of the XAUG + LRP approach is that all of these types of information can be used to gain an understanding of the relevant features. This work is similar but complementary to Ref.~\cite{Faucett:2020vbu}, which extracts expert information from the network for kinematic and shape variables, whereas we also investigate flavor information.

It is known that in many cases, complete decorrelation with kinematic variables (such as transverse momentum \pt and mass $m$) is desirable for many analyses (see an overview in Ref.~\cite{Larkoski:2017jix}). However, in the interest of simplicity, for this paper decorrelation is not addressed, although the general features of using XAUG variables and LRP extend to this case as will be demonstrated in future work.

In the following sections, we will introduce the layerwise relevance propagation technique (Sec.~\ref{sec:lrp}). We will then describe a toy model for demonstration in Sec.~\ref{sec:toy} to highlight how the technique works in a trivial but instructive case. Section~\ref{sec:particlemodel} will describe the particle-level model. Explanations of both the toy model and the particle model will be discussed in Sec.~\ref{sec:explanations}. Finally, we will present conclusions in Sec.~\ref{sec:conclusions}.

\section{Layerwise Relevance Propagation}
\label{sec:lrp}

LRP is a linearized approximation to networks that can be thought of as a "Deep Taylor decomposition"~\cite{Montavon_2017}. To understand this method, take a neural network with a prediction $f(x)$, based on some inputs $x$; these inputs can be pixels in an image or input variables, for instance. LRP propagates the prediction backwards through the network, eventually assigning a relevance score $R_j^{(1)}$ to each piece of input $x_j^{(1)}$, where $j$ enumerates the input features and the "1" in the superscript indicates the first layer, $L_{l}$, of the DNN. The relevance score indicates how much each input pixel or variable contributes to the final prediction.

In an ideal case, the LRP backwards propagation method has an overall relevance conservation at each layer $L_{l}$ as a sum over features $j$:

\begin{equation}
f(x)=\cdots=\sum_{j \in L_{l+1}} R_{j}^{(l+1)}=\sum_{j \in L_{l}} R_{j}^{(l)}=\cdots=\sum_{j} R_{j}^{(1)}.
\end{equation}
Again, the superscript indicates the layer the relevances are being calculated at, and the subscript indicates the summation over the relevances within that layer. Relevance conservation means that at every layer of the network, the total relevance sum is the same~\cite{Montavon2019}; and at each layer the relevance scores sum to the prediction. Therefore, the backwards propagation process does not alter the prediction, and LRP attributes the entirety of the network's decision to the inputs.

While there are many possible implementations of the LRP propagation rules ~\cite{Montavon2019}, we focus on only a few of them in this paper. First, we consider LRP-$\epsilon$ in which the relevance score for the $j$th neuron in layer $L_{l}$ is computed as:

\begin{equation}
    \sum\limits_{j}{R_j^{(l)}} = \sum\limits_{k}\frac{x_j w_{jk}}{\sum\limits_{i}{x_i w_{ik}}+\epsilon}R_k^{(l+1)}.
\end{equation}
Here, as in the other LRP rules, $x_j$ is the activation of the neurons at layer $l$, $R_k^{(l+1)}$ is the relevance scores assigned to the $L_{l+1}$ layer's neurons, and $w_{jk}$ is the weight connecting neurons $j$ and $k$~\cite{DBLP:journals/corr/abs-1708-08296}. The $\epsilon$ term is included for numerical stability. In LRP-$\epsilon$, two criteria determine how relevance is propagated from layer $L_{l+1}$ to each $L_{l}$ layer's neuron. The first criterion is $x_j$, the neuron activation. Rather intuitively, more relevance goes to more activated neurons. The second criterion is $w_{jk}$, where stronger connections (with larger $w_{jk}$) receive more relevance. The simple case where $\epsilon$ is set to zero is called LRP-z.

 In particular, for LRP-$\epsilon$, especially at large values, $\epsilon$ can absorb some of the relevance~\cite{10.1371/journal.pone.0130140}. Therefore, we also consider LRP-$\alpha \beta$:

\begin{equation}
    \sum\limits_{j}{R_j^{(l)}} = \sum\limits_{k}\left(\alpha  \frac{(x_j w_{jk})^+}{ \sum\limits_{i}{(x_i w_{ik})^+}} - \beta  \frac{(x_i w_{ik})^-}{ \sum\limits_{i}{(x_i w_{ik})^-}} \right)R_k^{(l+1)} ,
\end{equation}

where

\begin{minipage}{0.45\textwidth}
    $(x_j w_{jk})^+ = \begin{cases}
    x_j w_{jk} ;& x_j w_{jk} > 0\\
    0 ;& \mathrm{else}
    \end{cases}$
\end{minipage}
\hspace{-2mm} and \hspace{6mm}
\begin{minipage}{0.45\textwidth}
    $(x_j w_{jk})^- = \begin{cases}
    0 ;& \mathrm{else}\\
    x_j w_{jk} ;& x_j w_{jk} < 0
    \end{cases}$
\end{minipage} \\

indicate the positive and negative contributions to the relevance, respectively~\cite{DBLP:journals/corr/abs-1708-08296}. Positive contributions to the relevance correspond to a contributing activation function, and negative relevance corresponds to an inhibited activation function. Therefore, choosing different values of $\alpha$ and $\beta$ allows for control over the importance of positive and negative contributions to the network's decision~\cite{10.1371/journal.pone.0130140}. Relevance conservation is enforced by requiring $\alpha - \beta = 1$. 
In practice, ML models typically include a bias in each layer for stability; however, this violates relevance conservation. In such cases, the summation of all relevance scores does not equal the total score, although this is not a significant disadvantage.

In this paper, we implement LRP using the \textit{iNNvestigate} Python package~\cite{alber2018innvestigate}.
To compute the LRP score for CNN models, we use the "Preset A" mode of \textit{iNNvestigate}, which uses LRP-$\epsilon$ for dense layers, and LRP-$\alpha\beta$ for the convolution layers. To compute the LRP score for RNN models, we use LRP-$\epsilon$ throughout. These more practical LRP measures unfortunately do not conserve relevance at each layer, but this is not a major drawback. 

\begin{figure}[!ht]
    \centering
    \includegraphics[width = 0.4\textwidth]{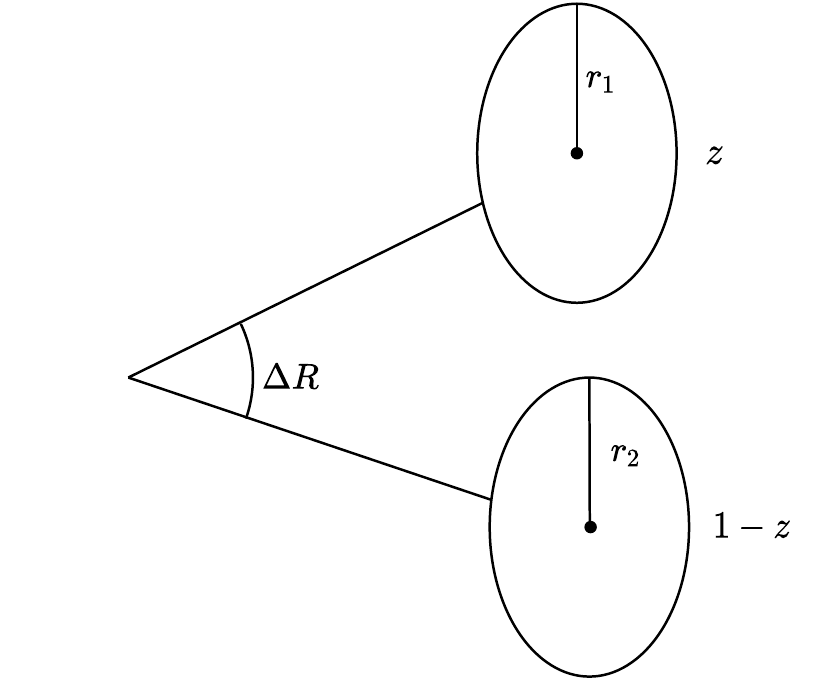}
    \includegraphics[width = 0.35\textwidth]{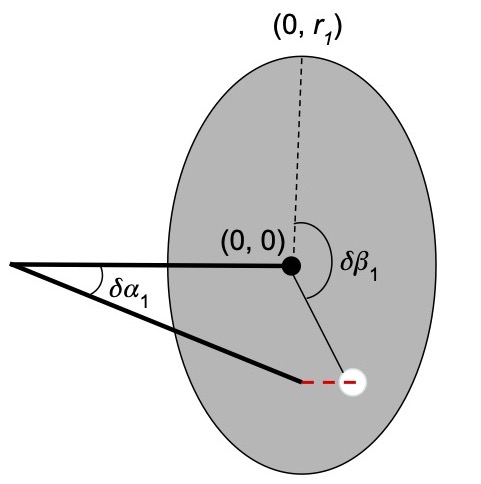}
    \caption{Diagram showing the toy "event" level parameters (left) and the toy "particle" level parameters (right).}
    \label{fig:alphaphi}
\end{figure}

\section{Toy model}
\label{sec:toy}

We first create a toy model to explore the feature exhaustion of networks using XAUG variables. The toy model is designed to vaguely mimic the salient features of jet images, to be processed by a 2D convolutional network, and a list of particle level information, to be processed by an RNN or 1D CNN. We assume there are two populations of inputs for each network, and design the networks to discriminate between the two populations. 

Each toy event corresponds to one jet consisting of 2 subjets, each comprised of 10 particles. For the toy images, the generated features are based on the substructure features of a jet with two subjets of radii $r_1$ and $r_2$, an angular distance of $\theta$ apart, with momentum fractions $z$ and $1-z$, as shown in Figure \ref{fig:alphaphi}. These parameters are cartoon-level models for a jet with two subjets, where $\theta$ corresponds to $\Delta R$ for a boosted jet with small angular separation as described in Ref.~\cite{Dasgupta:2013ihk}. The "jet momentum" is normalized to one, and since we are limiting the model to events with 2 subjets, the subjet momentum fractions are defined as $\z$ and $1-\z$. Using  these parameters, particles are randomly generated in each subjet; these are then pixelized into jet images. Each jet image is $16 \times 16$ pixels, representing calorimeter cells in a detector, with 20 particles in each image or event. 

\begin{figure}[!ht]
 \centering
        \includegraphics[width=0.32\linewidth]{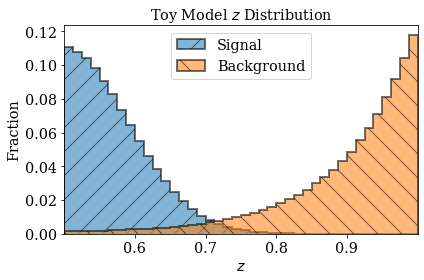}
        \includegraphics[width=0.32\linewidth]{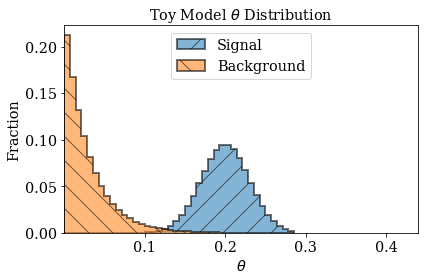}
        \includegraphics[width=0.32\linewidth]{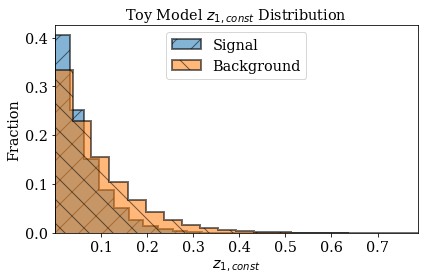}
        \includegraphics[width=0.32\linewidth]{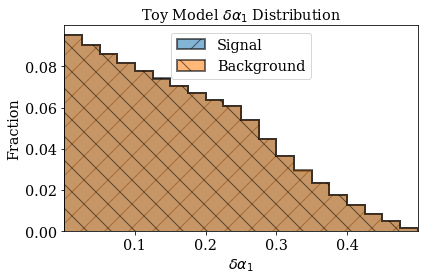}
        \includegraphics[width=0.32\linewidth]{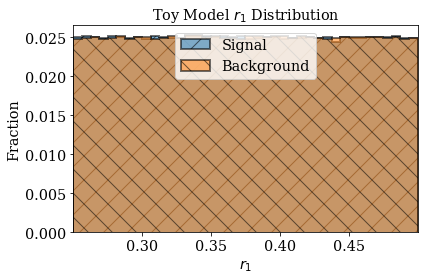}
        \includegraphics[width=0.32\linewidth]{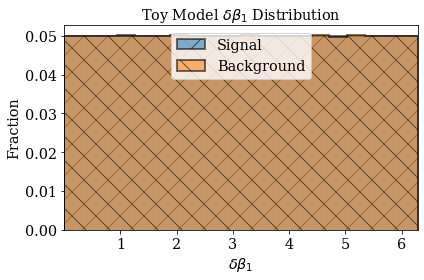}
    \caption{The normalized $\boldmath{z}$, $\theta$ distributions for signal and background events in the Toy Model, along with the $z_1$, $\delta\alpha_1$, $r_1$, and $\delta\beta_1$ values for those same events.}
    \label{fig:toyinputdist}
\end{figure}

\begin{figure}[!htb]
 \centering
    \begin{subfigure}{.47\textwidth}
        \includegraphics[width=\linewidth]{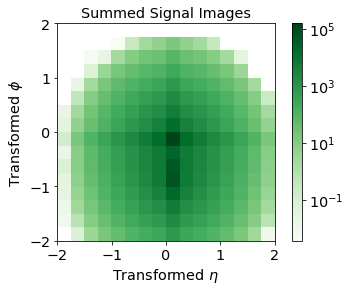}
        \caption{}
        \label{fig:toysig}
    \end{subfigure}
    \begin{subfigure}{.47\textwidth}
        \includegraphics[width=\linewidth]{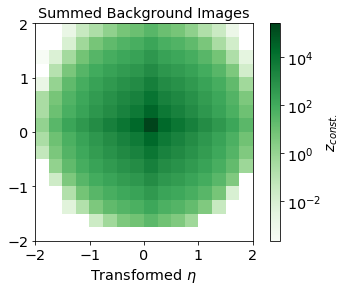}
        \caption{}
        \label{fig:toybg}
    \end{subfigure}
    \caption{Toy model summed input images for signal (a) and background (b) jets.}
    \label{fig:toyimages}
\end{figure}

 There are two image populations, which we will refer to as \textbf{signal} and \textbf{background}. We generate 1M images, half signal and half background.
 The signal is randomly sampled from normal distributions
 for \z and $\theta$, 
 while the background is sampled from exponential distributions that are mostly 
 separated from signal in their combined phase space (see Table \ref{tab:toytable}). The radii of the subjets for both cases are sampled from 
 a uniform distribution in a range that keeps the particles within the $\mathbf{16 \times 16}$
 image. These radii enclose the sampling of the particles coordinates with respect to the 
 subjet axis, which are represented by variables $\delta\alpha$ and $\delta\beta$.
 The angular distance from the subjet axis, $\delta\alpha$
 is sampled from an additional exponential  distribution with a scale parameter of $0.5$ for both signal and background, limited by the radius of the subjet. The azimuthal angle with respect to the subjet axis, $\delta\beta$, is sampled from a uniform distribution of $[0, 2\pi)$ for both populations. The momentum fraction of the subjet \z is distributed among the 10 particles of the subjet according to the distributions, denoted $z_1$ for the leading subjet and constrained to sum to \z, and $z_2$ for the second subjet, constrained to sum to $1-z$. These can be interpreted as the momenta of the constituents, and are used as the pixel intensities in the jet image. We obtain the coordinates of the pixels within the image by transforming $\delta\alpha$ and $\delta\beta$ to an abstracted $\eta$-$\phi$ plane via a simple rotation. In the $\eta$-$\phi$ plane, the leading-\pt subjet axis is at $\boldmath(0,0)$ and the subleading-\pt subjet axis is at $\boldmath(0,-1)$. Finally, the image is flipped if the sum of pixel intensities in the left half of the image is greater than that on the right half.  The input distributions are shown in Figure \ref{fig:toyinputdist}, and examples of summed input images are shown in Figure \ref{fig:toyimages}.

\begin{table}[!htb]
\centering
\begin{tabular}{llll}
\hline
& \multicolumn{2}{l}{Signal} & Background   \\ \hline & $\mu$ & $\sigma$ & $\beta$ \\ \hline
\multicolumn{1}{|l|}{\z}     & \multicolumn{1}{l|}{0.2} & \multicolumn{1}{l|}{0.03} & \multicolumn{1}{l|}{0.03} \\ \hline
\multicolumn{1}{|l|}{$\theta$} & \multicolumn{1}{l|}{0.5} & \multicolumn{1}{l|}{0.09} & \multicolumn{1}{l|}{0.01} \\ \hline
\end{tabular}
\caption{Parameters used to generate toy model distributions. $\mu$ and $\sigma$ are the parameters for a gaussian distribution and $\beta$ is the parameter for an exponential of the form $f(x) = \frac{1}{\beta}\exp(\frac{-x}{\beta})$.}
\label{tab:toytable}
\end{table}
 
 For use in networks that take 1-dimensional inputs, the data is structured as a list of the simulated "particle-level" information, i.e., the individual constituent $\z$ fractions, and $\delta\alpha$ and $\delta\beta$ values, as seen in the right diagram of Figure \ref{fig:alphaphi}. Within each event, particles are sorted by $\z_1$, which is the leading subjet's $\z$ value distributed among that subjet's particles. We refer to this as a "particle list", and just like the image data, we generate 1M events, half signal and half background.

We investigate three network structures: a 2D CNN for jet images, a 1D CNN for particle lists, and a 1D RNN for particle lists. 
The datasets were shuffled and split into testing, training, and validation subsets. In all cases, when XAUG variables are used in conjunction with low-level inputs, they are concatenated to the post-processed outputs of the CNN or RNN. This augmented list is then combined in a flattened layer for input to an MLP that provides the final decision. 

\subsection{2D CNN for toy model}
\begin{figure}[!ht]
    \centering
    \includegraphics[width=\textwidth]{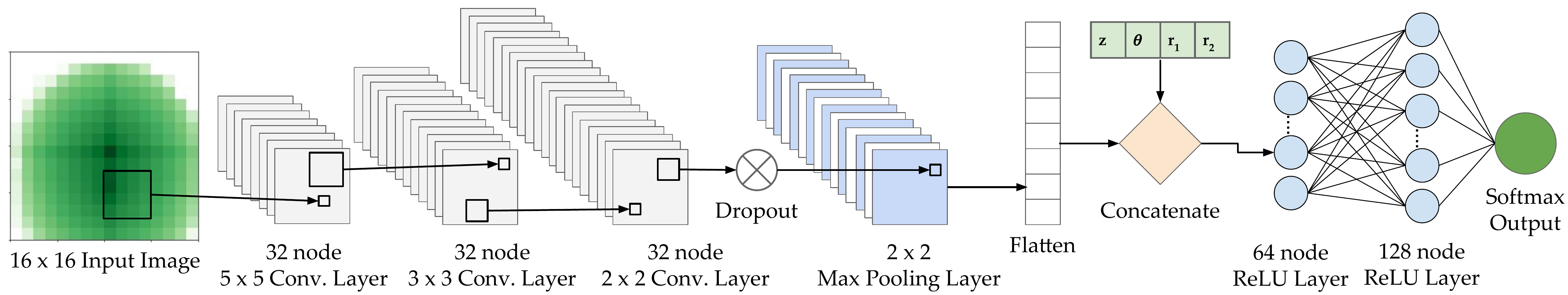}
    \caption{Diagram of 2D CNN for classification of toy model images. When used, XAUG variables are concatenated to the processed outputs immediately before the final two ReLU layers.}
    \label{fig:toyCNN2D}
\end{figure}

For the 2D CNN, we build a network consisting of 3 convolutional layers, followed by a dropout layer that randomly drops 20\% of the nodes, and a max-pooling layer that calculates the maximum value of the feature map, then 2 dense rectified linear unit (ReLU~\cite{ReLU_1, ReLU_2}) layers, with a softmax dense layer making the final classification between the signal and background populations. This network is based on networks made to analyze 2D jet images, like those in Refs.~\cite{Cogan:2014oua,deOliveira:2015xxd}. The network that operates on the toy images has 154,178 trainable parameters. We concatenate the "event" level XAUG variables to the end of the flattened convolutional output which is then processed through two dense layers to further find relations in the concatenated input.
The final network has 43,073 trainable parameters, the details of which can be seen in Fig.~\ref{fig:toyCNN2D}. 

\subsection{1D CNN for toy model}

\begin{figure}[!ht]
    \centering
    \includegraphics[width=1.1\textwidth]{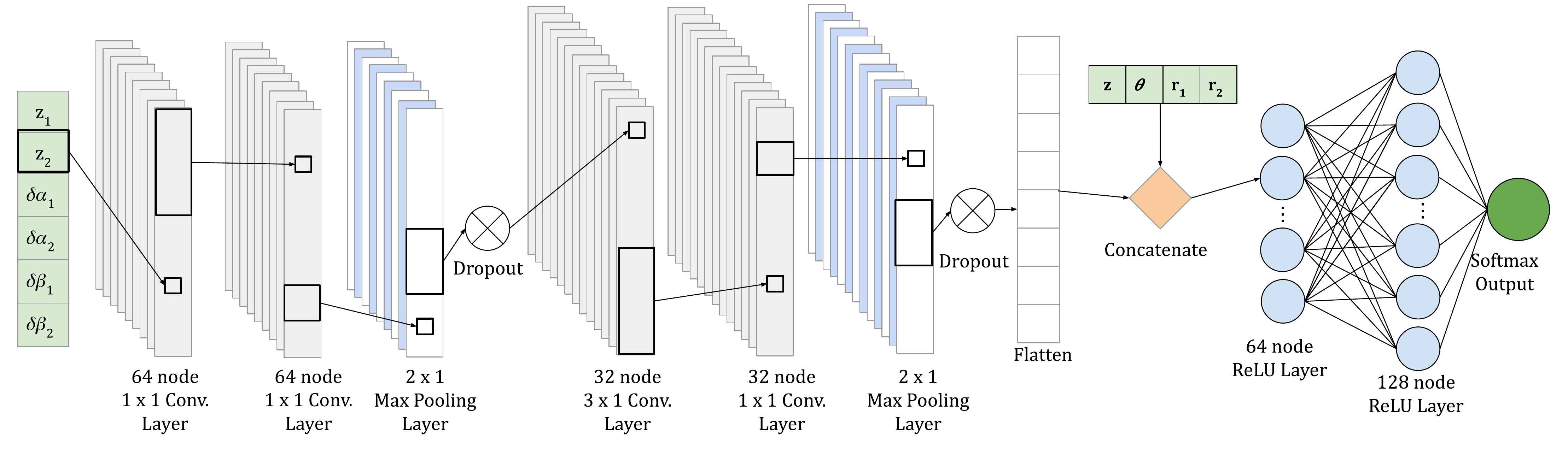}
    \caption{Diagram of 1D CNN for classification of toy model particle lists. When used, XAUG variables are concatenated to the processed outputs immediately before the final two ReLU layers.}
    \label{fig:toyCNN1D}
\end{figure}

For the 1D CNN, we build a network consisting of a set of layers that goes over each set of information within the particle list separately before the outputs of these layers are concatenated together and passed through a dense layer before a final decision is made. Each set of shape (20, 1) input tensors goes through two 1-dimensional convolutional layers of stride 1 - one with kernel size 3 followed by a layer with kernel size 1 - before being fed to a dropout layer that drops 20\% of the nodes, and a max pooling layer. This set of layers is repeated, except the convolutional layers not have half the nodes of their previous iteration, before the tensors are flattened and concatenated with one another. In the case where XAUG variables are added, they are concatenated with the particle level information after they've gone through the aforementioned layers. This model has 270,082 trainable parameters, the details of which can be seen in Fig.~\ref{fig:toyCNN1D}.  

\subsection{1D RNN for toy model}

For the toy model RNN, we build a network consisting of recurrent layers that process the particle level information before being flattened and passed through a dense layer before the final decision is made. The inputs - which are a size (10, 6) tensor containing the \z, $\delta\alpha$, and $\delta\beta$ information for the constituents of the subjets and size (1,) tensors for the XAUG variables - are divided into subsets based on the chosen timestep of the recurrent layers. In our network there are 10 timesteps per event; this equates to processing information from 2 particles at a time, with a memory of the previous 2 particles being used to find patterns within the event. We found this method to be the most performant and efficient for our network, compared to e.g. processing one particle or all particles at a time.

\begin{figure}[!ht]
    \centering
    \includegraphics[width=\textwidth]{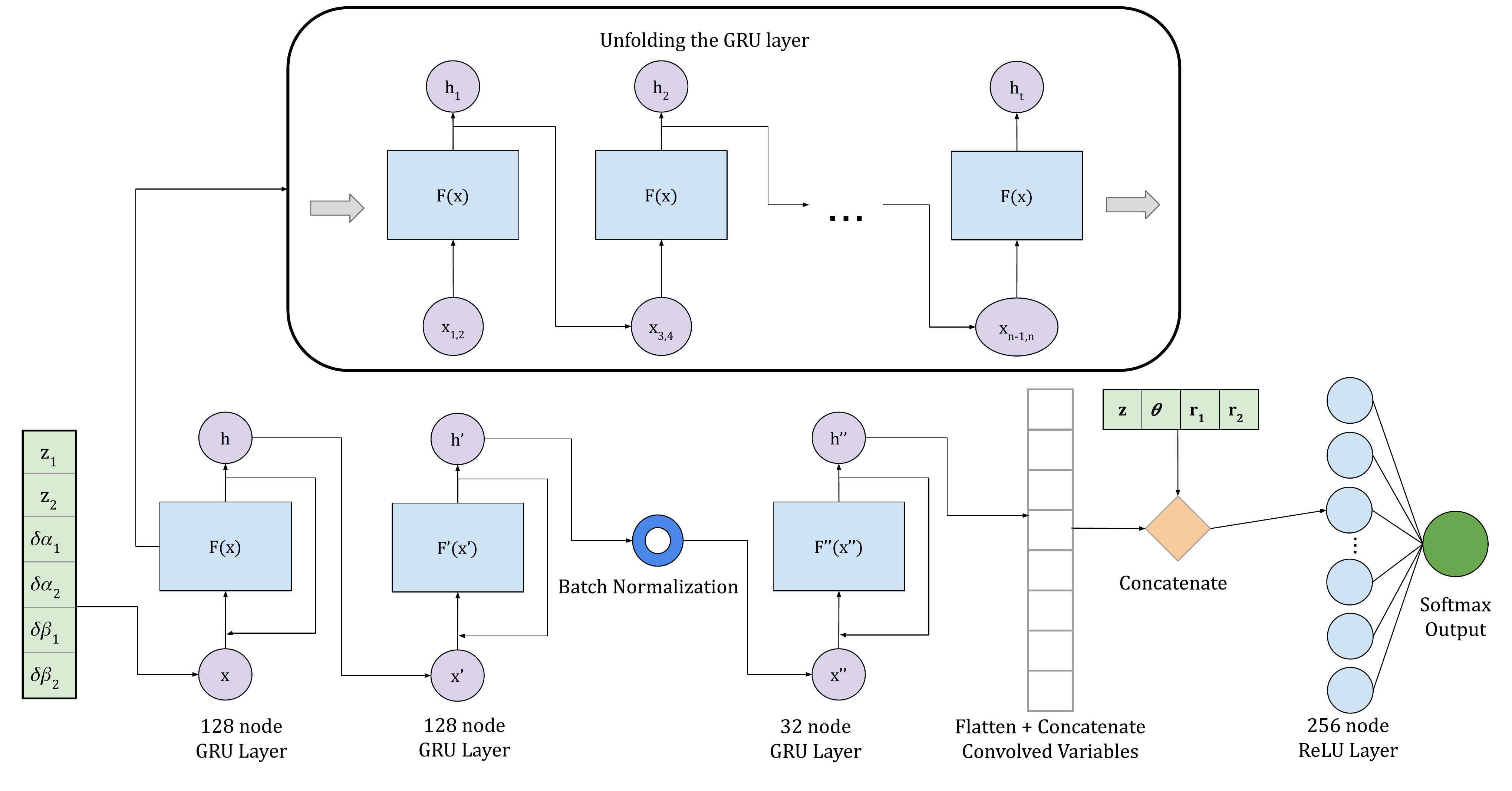}
    \caption{Diagram of RNN for classification of toy model particle lists. When used, XAUG variables are concatenated to the processed outputs immediately before the final dense layer.}
    \label{fig:toyRNN}
\end{figure}

The particle list information passes through 2 gated recurrent unit layers, then a batch normalization, one more gated recurrent unit, before being flattened and passed to the dense layers for decision-making. In the case where XAUG variables are added, they are concatenated with the particle level information after they've gone through their respective recurrent layers. This model has 248,938 trainable parameters, and the details can be seen in Fig.~\ref{fig:toyRNN}.

\section{Particle-level model}
\label{sec:particlemodel}

The more realistic simulation is a set of {\tt PYTHIA} 8.2.35~\cite{Sjostrand:2014zea} events.\footnote{Our software is outlined \href{https://github.com/rappoccio/PythiaGenJets}{here}} To simulate the 2-prong structure of boosted jets, we produce SM $ZZ$ production with both $Z$ bosons decaying to $\bbbar$ \footnote{The configuration for SM $ZZ$ production is \href{https://github.com/rappoccio/PythiaGenJets/blob/master/zz_bb_flatter.cfg}{here}}. Both $b$ quarks from the $Z$ decay are required to be within $\Delta R < 0.8$ of the $Z$. We use generic QCD multijet events as background\footnote{The configuration for QCD multijet event simulation is \href{https://github.com/rappoccio/PythiaGenJets/blob/master/qcd_flat15to7000.cfg}{here}}. We only consider the leading jet in the simulations, so we refer to these two samples as $Z{\bbbar}$ and QCD, respectively. 

\begin{figure}[!ht]
    \begin{subfigure}[t]{0.32\textwidth}
    \centering
    \includegraphics[width=0.9\linewidth]{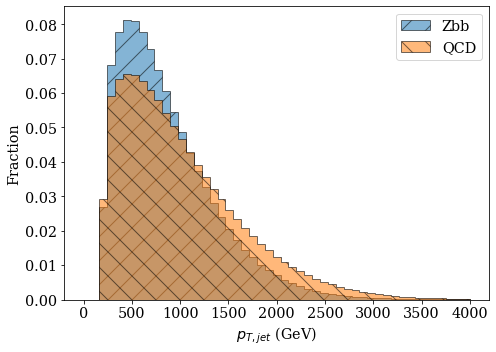}
    \caption{AK8 jet \pt}
    \label{fig:jetpt}
    \end{subfigure} 
    \begin{subfigure}[t]{0.32\textwidth}
    \centering
    \includegraphics[width=0.9\linewidth]{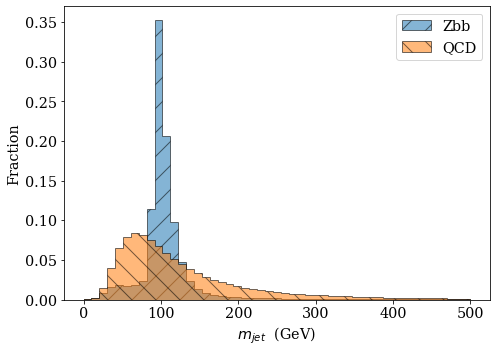}
    \caption{AK8 jet mass}
    \label{fig:jetMass}
    \end{subfigure}
    \begin{subfigure}[t]{0.32\textwidth}
    \centering
    \includegraphics[width=0.9\linewidth]{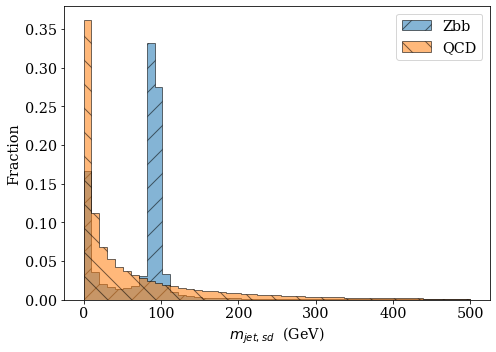}
    \caption{AK8 soft dropped jet mass}
    \label{fig:jetMassSD}
    \end{subfigure}

    \begin{subfigure}[t]{0.32\textwidth}
    \centering
    \includegraphics[width=0.9\linewidth]{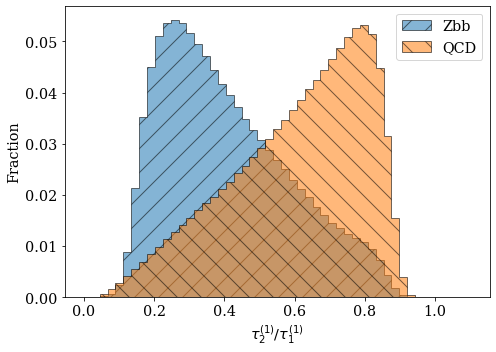}
    \caption{$\tau_2 / \tau_1$}
    \label{fig:jettau21}
    \end{subfigure}
    \begin{subfigure}[t]{0.32\textwidth}
    \centering
    \includegraphics[width=0.9\linewidth]{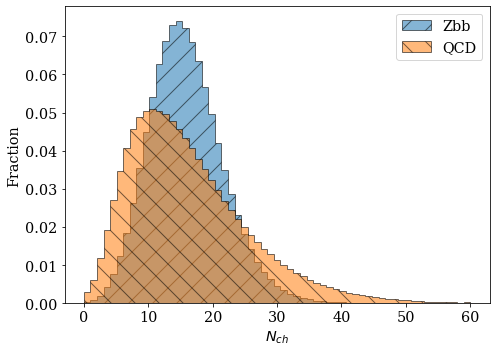}
    \caption{Number of charged hadrons}
    \label{fig:n_ch}
    \end{subfigure}
    \begin{subfigure}[t]{0.32\textwidth}
    \centering
    \includegraphics[width=0.9\linewidth]{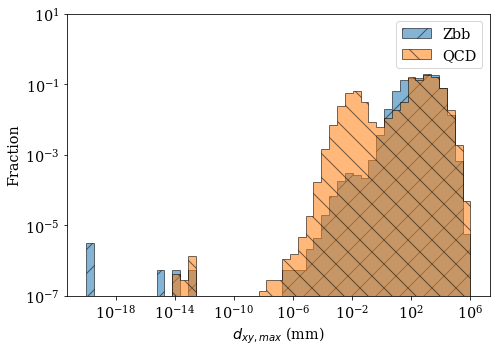}
    \caption{Max $d_{xy}$ per event}
    \label{fig:dxy_max}
    \end{subfigure}

    \caption{Input distributions to training for $Z\bbbar$ and QCD events. Shown are the jet $\pt$ (upper left), ungroomed jet mass (upper middle), groomed jet mass with the soft drop algorithm (upper right), n-subjettiness ratio $\tau_2/\tau_1$ (lower left),  charged hadron multiplicity (lower middle), and $d_{xy,\ max}$ (lower right).
    \label{fig:inputdists}
    }
\end{figure}

It has been shown in Ref.~\cite{Dolen:2016kst} that independence of kinematic variables is a desirable feature for boosted object tagging, and has been applied in several "decorrelated" versions of DNNs, with examples in Refs.~\cite{Shimmin:2017mfk,CMS:2019gpd,Kitouni:2020xgb}. Examples of approaches taken in other works have been to train an adversarial network to achieve the desired independence from kinematic variables, reweighting the samples, or to decorrelate the behavior in a brute force approach. Our simulated samples have different $\pt$ spectra in reality, however we wish to eliminate such kinematic differences to some extent in our training. As such, we artificially weight the {\tt PYTHIA} generation with the {\tt bias2Selection} parameter in order to have similar jet \pt spectra, which is shown in Fig.~\ref{fig:jetpt}.
We therefore partially remove the \pt and mass dependencies of the classifier. The goal of this paper is to explain decisions, so we take this simpler approach for ease of demonstration. It does not achieve complete decorrelation. The principles we develop are also applicable to more thoroughly decorrelated taggers, but become more complicated due to the increased dimensionality of having adversarial networks and are thus left to future work for investigation.

We then cluster the particles with {\tt fastjet} 3.3.1~\cite{Cacciari:2005hq,Cacciari:2011ma} to create anti-$\kt$ jets~\cite{Cacciari:2008gp} with a distance parameter of $R=0.8$. We then use the {\tt RecursiveTools} package from {\tt fastjet-contrib} 1.036, which implements the modified Mass Drop Tagger (mMDT)~\cite{Dasgupta:2013ihk}. This is the same algorithm as the Soft Drop (SD) algorithm~\cite{Larkoski:2014wba} with $\beta=0$. We use the mMDT with $z_{cut}=0.1$, $\beta=0$ and a cut to require exactly two subjets within every jet. 
We define the groomed or soft dropped jet mass as $m_{SD}$. 

We select the leading jet in the event with $\pt > 200$ GeV, $|\eta| < 2.4$, and $50 < m_{SD} < 150$ GeV. After these selections, there are 552k events in the signal and background categories (half in each). The distributions for the jet $\pt$, mass, and soft dropped mass are shown in Fig.~\ref{fig:inputdists}. 

The information used for each particle in our classifiers are the four-vector information ($\pt$, $\eta$, $\phi$, mass), the production vertex spatial location ($d_x$, $d_y$, $d_z$, with transverse distance denoted by $d_{xy}$), and the particle PDG ID \cite{PhysRevD.98.030001}. In addition, the distance to the center of the jet, and the distances to each subjet are also stored. The distribution for $d_{xy}$ is shown in Fig.~\ref{fig:inputdists}. 

\subsection{XAUG variables for particle-level model}

To encapsulate the kinematic information of jets and their substructure, we will select the $N$-subjettiness variables $\tau_{m}^\beta$ as one set of XAUG variables. We also input the jet four-vector, and the groomed jet mass.\footnote{Most taggers have historically avoided adding the mass to the classifier inputs, however in our case we have decided to add it since it can be deduced from the $N$-subjettiness variables in any case.} The angular distance between the two subjets $\Delta R_{\mathrm{subjets}}$ and momentum fraction of the leading subjet $z$ are also added. 

To add simple flavor identification, we add jet composition variables such as charged hadron multiplicity ($N_{\mathrm{ch}}$), neutral hadron multiplicity($N_{\mathrm{neut}}$), photon multiplicity ($N_{\gamma}$), muon multiplicity ($N_{\mu}$), and electron multiplicity ($N_{e}$). A measure of the soft radiation in the jet is represented by the jet pull angle ($\phi_{\mathrm{pull}}$)~\cite{Gallicchio:2010sw}. Finally, heavy flavor identification information is encapsulated by an (over)simplified metric of the maximum of the spatial locations of the production vertex for each particle in the $x-y$ ($d_{xy}$) and $z$ ($d_{z}$) directions. Distributions for $\tau_{2}/\tau_{1}$ and $N_{\mathrm{ch}}$ are shown in Fig.~\ref{fig:inputdists}. 

\subsection{Variable normalization for particle-level model}

To ensure that all of the inputs to our networks have comparable numerical magnitude, we perform preprocessing on each input variable based on the overall distribution in the signal and background. An equal number of signal and background events are chosen in each case, and then the distribution of each input variable for the ensemble are calculated. The mean and standard deviations are computed, and the distribution is truncated between $\pm 3$ standard deviations of the mean. We then define the minimum to be zero, and the maximum to be one to remove network bias towards inputs with very large values. These are referred to as "normalized" variables. To clarify, the underflows and overflows appear at 0 and 1, respectively.


\subsection{Classifiers for particle-level model}

We consider three types of classifiers. The first is a 2D image-based CNN, with preprocessing inspired by Refs.~\cite{Cogan:2014oua,deOliveira:2015xxd}. The second is a 1D CNN that inputs measurements from the first $N$ particles in a transverse momentum (\pt) ordered list, similar to the {\tt DeepAK8} algorithm in Ref.~\cite{CMS:2019gpd}. The third is a recurrent neural network (RNN) that inputs an arbitrarily large list of the same information as the 1D CNN.

We discuss each classifier in turn. 

\subsection{2D CNN for particle-level model}
The first classifier we consider for particle-level simulations is a CNN that is based on 2D jet images, with preprocessing inspired by Refs.~\cite{Cogan:2014oua,deOliveira:2015xxd}. We then use a CNN with the same architecture shown in Fig.~\ref{fig:toyCNN2D}, but with different XAUG variables and the full list of particle inputs. At this point, if only looking at image data, the network will pass the flattened images to 2 dense layers before being output. If the network is also passed XAUG variables, these will be concatenated with the flattened images before going through the dense and output layers.

\begin{figure}[!ht]
    \centering
    \includegraphics[width=1.1\textwidth]{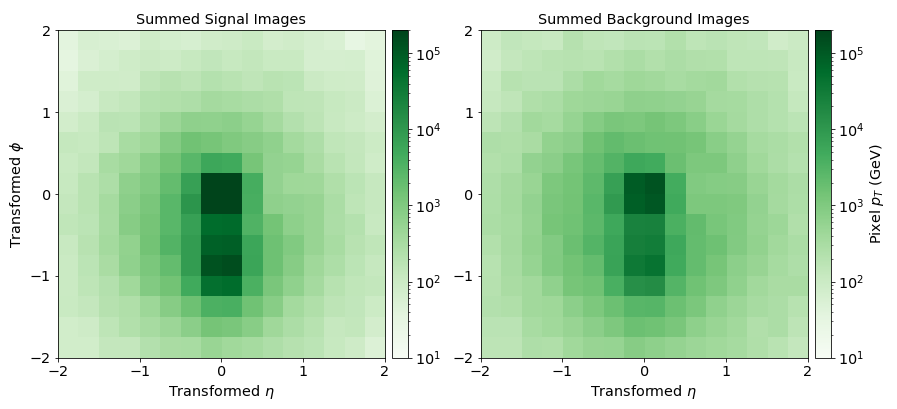}
    \caption{Aggregated Signal ($Z\bbbar$) and Background (QCD) Jet Images after preprocessing.}
    \label{fig:jet2DCNN-images}
\end{figure}

The preprocessing is as follows. First, we find exactly two subjets (events with fewer than two subjets are discarded). We then examine the leading jet in the event, and calculate the subjet directions. The subjet with the higher $\pt$ is used as the center of the jet image $(0,0)$, and the image is then rotated in local pseudorapidity-azimuth space ($\eta-\phi$) such that the lower-$\pt$ subjet is pointing downward. The radial distances are scaled in units of the $\Delta R$ between the two subjets, such that the lower-$\pt$ subjet is placed at $(0,-1)$. The $\pt$ of all of the constituents are scaled by the total jet $\pt$, and are then pixelated into a grid of size $\mathbf{16 \times 16}$. The image is then parity flipped so that the largest sum of the pixel intensities is on the right-hand side of the image. Plots of the aggregated Signal ($Z\bbbar$) and background (QCD) are shown in Fig.~\ref{fig:jet2DCNN-images}. 


\subsection{1D CNN for particle-level model}

The second classifier we considered for particle-level simulations is a CNN that is based on observables from the first $N$ particles in a $\pt$-ordered list of inputs, similar to the {\tt DeepAK8} algorithm in Ref.~\cite{CMS:2019gpd}. One of the main advantages of {\tt DeepAK8} is the use of particle-level information to have sensitivity to particle content such as hadron flavor, and quark-gluon discrimination. For this reason, we have two separate particle list taggers, one with only the four-momentum information of the constituents, and another that adds other observables. We are interested in general characteristics of a procedure in this paper, so we do not attempt a realistic flavor tagging for our simulated sample. Instead, we directly input the particle ID of the stable particle, and the $x,y,z$ position of its production vertex. These are overly simplistic assumptions for a realistic tagger, however it will demonstrate the procedure of how flavor information could be investigated with our method. 

In particular, we have found that $N = 20$ provides good discrimination between 2-prong jets and QCD jets. We use a network architecture of 2 convolutional layers, one with shape (64, 3) and one with shape (64, 1) followed by a max pooling with pool size 2, followed by a dropout of $20\%$ of the network's nodes. The inputs tensors have shape (20, 1) and contain the particle list information of the leading constituents of the leading jet, sorted by $\pt$. The two convolutional layers and max pooling layer are repeated with shapes (32, 3), (32, 1) and 2 respectively with another $20\%$ dropout following. The convolved values are then flattened and concatenated with the XAUG input features, which are then fed through a dense layer with a ReLU activation. The network's structure is the same as in Fig.~\ref{fig:toyCNN1D} with different XAUG variables.

For inputs to the 1D CNN that are similar to those in {\tt DeepAK8}, 17 variables were created. The variables are listed in Table~\ref{tab:vars}.

\begin{table}[!ht]
\centering
\begin{tabular}{|c|}
\hline
Variable \\\hline
$\log(\pt)$  \\
$\log(\pt/{\mathrm {jet} \; \pt})$   \\
$\mathrm{log}(E)$ \\
$|\eta|$  \\
$\Delta \phi(\mathrm{jet}) $\\
$\Delta \eta(\mathrm{jet}) $\\
$\Delta R(\mathrm{jet}) $\\
$\Delta R(\mathrm{subjet_{1}}) $\\
$\Delta R(\mathrm{subjet_{2}}) $\\
Charge $q$ \\
isMuon \\
isElectron \\
isPhoton \\
isChargedHadron \\
isNeutralHadron \\
$d_{xy}$ \\
$d_{z}$ \\
\hline
\end{tabular}
 \caption{\label{tab:vars} Particle list input variables of 1D CNN, properties of the constituents of the leading jet.}
\end{table}

\subsection{1D RNN for particle-level model}

The final classifier is an RNN that uses similar inputs as the CNN from the particle list; however, we limit the number of particle and expert variables due to the increased training and processing time of the network. The network architecture we apply is a first input layer, followed by 3 recurrent layers, 2 additional input layers, and 2 dense layers at the end, shown in Fig.~\ref{fig:toyRNN}. 

For the first input, the network takes in the four-vector ($p_T$, $\eta$, and $\phi$) data of the first 20 constituents of the leading jet of each event sorted by $\pt$. This is given to the recurrent layers. Then, the XAUG variables $\tau_{21}$ and charged-hadron multiplicity are added before the final dense layers.

\section{Explanations}
\label{sec:explanations}

In the following section, we will discuss the performance and explanations of both the toy and particle models described in the previous sections. 

\subsection{Toy Model Explanations}

The classification of the toy signal and background is trivial, even without the XAUG variables. The area under curve (AUC) of the ROC curve is close to unity. However, it is still instructive to investigate the explanations of these decisions to demonstrate the features of LRP and XAUG variables. The understanding gained can also be applied to the more realistic particle simulation.

\begin{figure}[!ht]
 \centering
    \includegraphics[width=\linewidth]{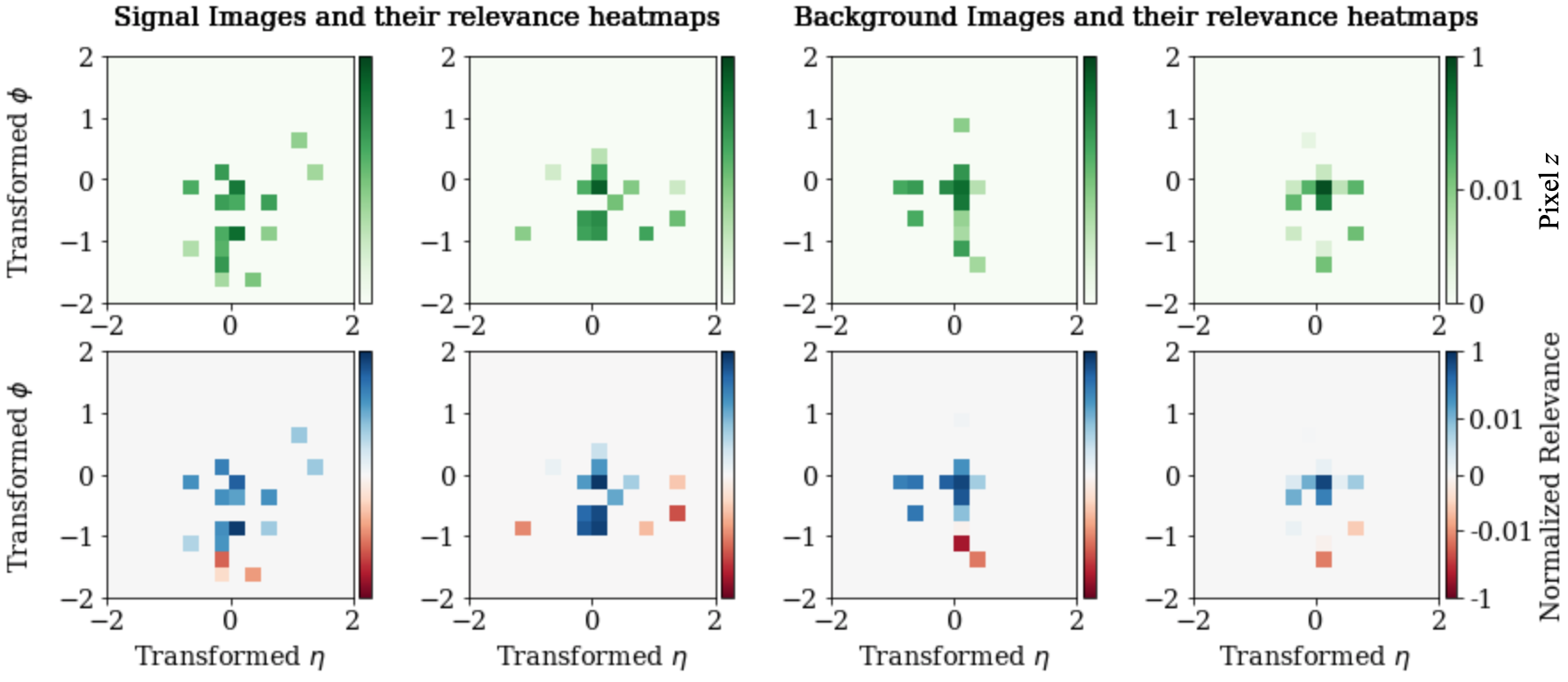}
    \caption{Individual Toy Model signal (left two columns) and background (right two columns) that were correctly predicted (top) and their corresponding LRP score heatmaps (bottom). Features with positive relevance support the prediction for the event, whereas features with negative relevance oppose the prediction for the event. Images are plotted in transformed pseudorapidity-azimuth ($\eta-\phi$) space.}
    \label{fig:toyimageslrp_individual}
\end{figure}

Examples of the classification in the 2D CNN are shown in Fig.~\ref{fig:toyimageslrp_individual}. There are four examples of toy signal and background events in green. Their LRP scores are also shown as heatmaps in red and blue. The blue LRP scores indicate pixels that contributed positively to the identification, whereas the red LRP scores indicate pixels that contributed negatively to the identification. It is clear that the signal classifications prefer information where there is a disjoint subjet near \textbf{(0,-1)}, whereas background classifications discourage that area. Indeed, the input signal images in Fig.~\ref{fig:toyimages} have a well-separated second subjet near \textbf{(0,-1)}, while the background images do not. As such, the network is clearly learning the appropriate information that was used to create the Toy Model, including the position (i.e. angle) and intensity (i.e. energy) of the "subjets". 

We then trained a network with the XAUG variables $r_1$, $r_2$, $z$, 
and $\theta$ included among the network input features; these variables are defined and plotted in Figs.~\ref{fig:alphaphi}~and~\ref{fig:toyinputdist}, respectively.
This is an exhaustive list of the entire information content of the toy model. As such, we expect that the network will be able to (almost) entirely ignore the information contained in the jet images, because its decision is completely determined by the XAUG variables. Any residual relevance is an artifact of the optimization, not a distinguishing feature. 

\begin{figure}[!ht]
    \centering
    \includegraphics[width=0.99\textwidth]{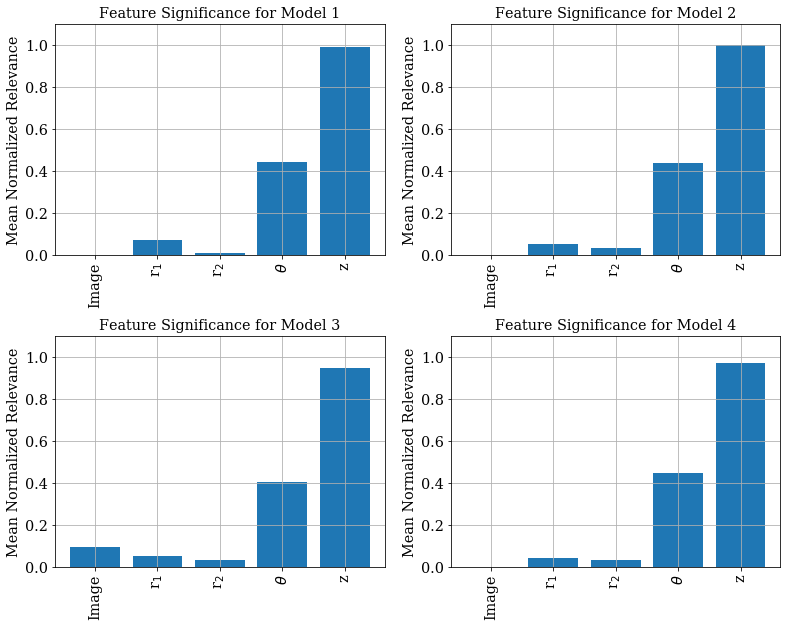}
    \caption{LRP scores for all variables in the Toy 2D CNN Model, with 4 separate trained models.}
    \label{fig:Toy_XAUG_featureSignificance_multimodel}
\end{figure}

To investigate the optimization dependence, we trained the same 2D CNN with images and XAUG variables four times, labeled Models 1-4. This allows us to have four separate optimizations and judge their consistency. The mean normalized LRP scores of the aggregated events are shown in  Fig.~\ref{fig:Toy_XAUG_featureSignificance_multimodel}. To produce these bar plots, for each event we first found the feature (XAUG or image pixel) with the maximum absolute LRP score. Then, in order to eliminate the dependence of the relevance on the events with large deviations in total relevance, we normalized the event by dividing all LRP scores by this maximum value. For each event, we then summed the absolute values of the normalized pixel LRP scores to get an image LRP score. After this, we averaged the absolute values of the XAUG and image LRP scores across all events to produced the mean normalized relevances shown. This procedure is used for all bar plots.

It is clear from these bar plots that the two most important pieces of information are the XAUG variables $z$ and $\theta$. There is very little information that the network learns from the "subjet" radii (which is expected, since they are drawn from the same uniform distributions). Furthermore, the network does not extract many features from the image itself after the addition of the XAUG variables. The image relevance varies between 0.0 to 0.1, but is always much lower than that of $z$ and $\theta$. This indicates an extremely flat optimization in the network training in the space of the input image. 
The network makes use of subleading features, although the extent varies if the network is retrained.

\begin{figure}[!ht]
    \centering
    \includegraphics[width=0.99\textwidth]{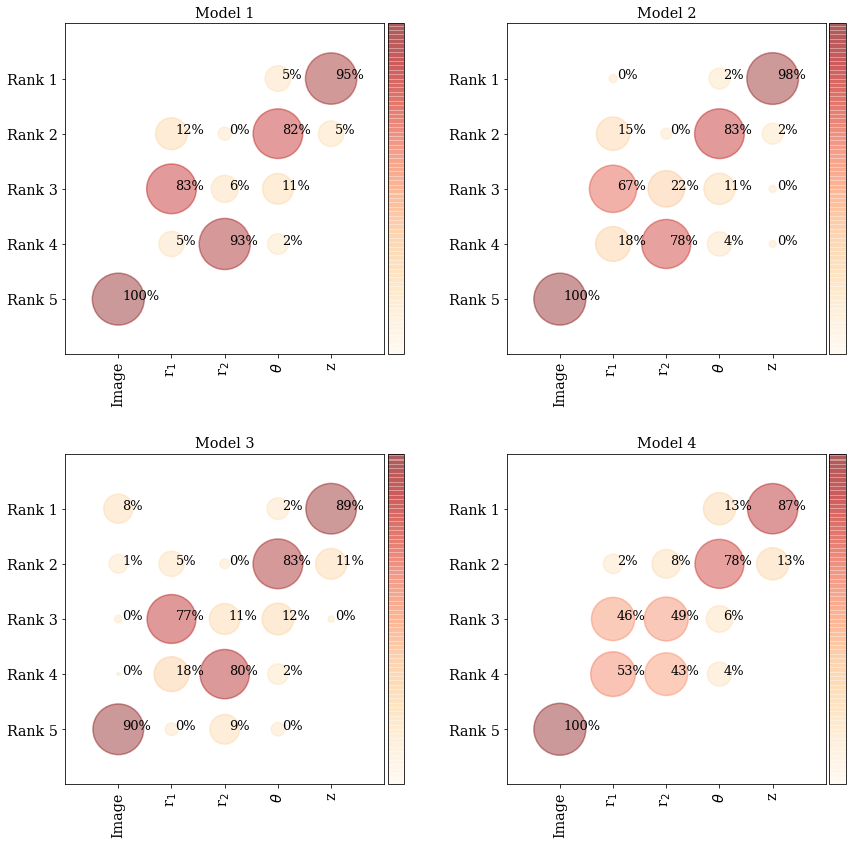}
    \caption{Ranked LRP scores for all variables in the Toy 2D CNN Model shown here for four separately trained models.}
    \label{fig:Toy_XAUG_z_vs_theta_predict_models_ranked}
\end{figure}

To study the impact of various features on the decision making, we plot the relative rank of each feature in Fig.~\ref{fig:Toy_XAUG_z_vs_theta_predict_models_ranked}. The annotation on each bubble is the percent of events for which the variable had a certain rank. The $z$ variable is ordinarily the highest ranked in all of the four models we used, while the image is ordinarily last. The variables $r_1$ and $r_2$ do not carry any discriminatory powers, as can be seen from Fig.~\ref{fig:Toy_XAUG_featureSignificance_multimodel}.  

It is interesting to note that there is some variation within the optimizations, especially in the subleading domain of the optimization space. Differences in relevance attributed to features between trainings point toward differences in local optimization of the network. 

\begin{figure}[!ht]
    \centering
    \includegraphics[width=0.99\textwidth]{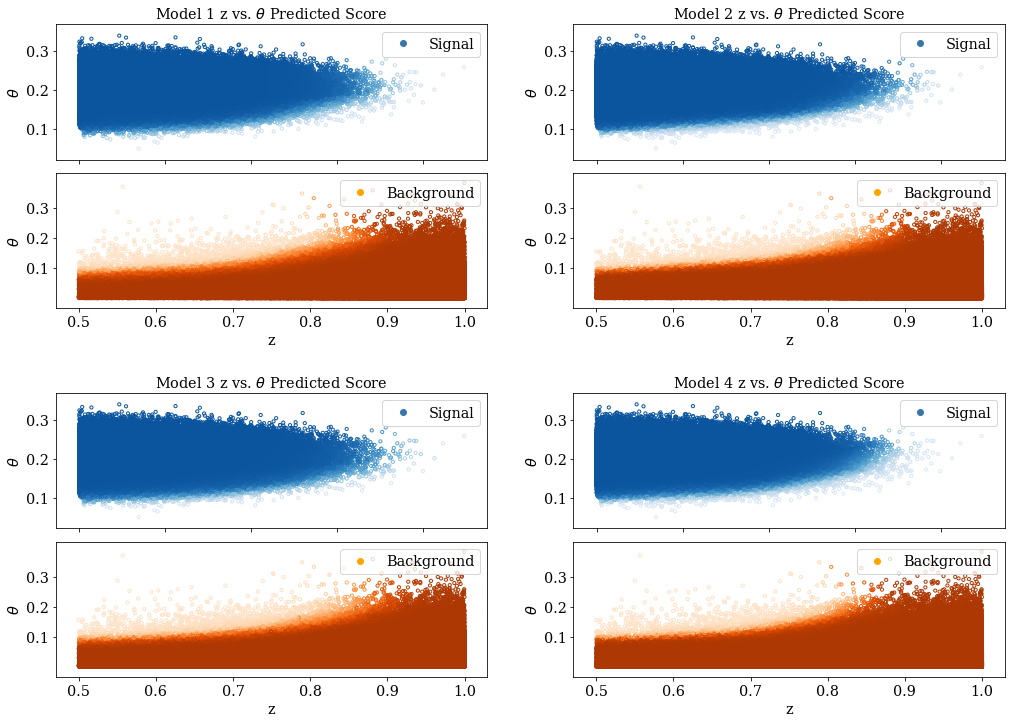}
    \caption{Predicted score for Toy Model for four separately trained models. 
    The momentum fraction $z$ is shown on the $x-$axis while the angle $\theta$ is shown on the $y-$axis. 
    The orange distribution is the background and the blue distribution is the signal. The intensity of the color indicates the predicted scores. While the overall classification boundary is the same for the four trainings, we can see subtle differences in the prediction boundary when comparing different trainings.}
    \label{fig:Toy_XAUG_z_vs_theta_predict_model1}
\end{figure}

\begin{figure}[!ht]
    \centering
    \includegraphics[width=0.99\textwidth]{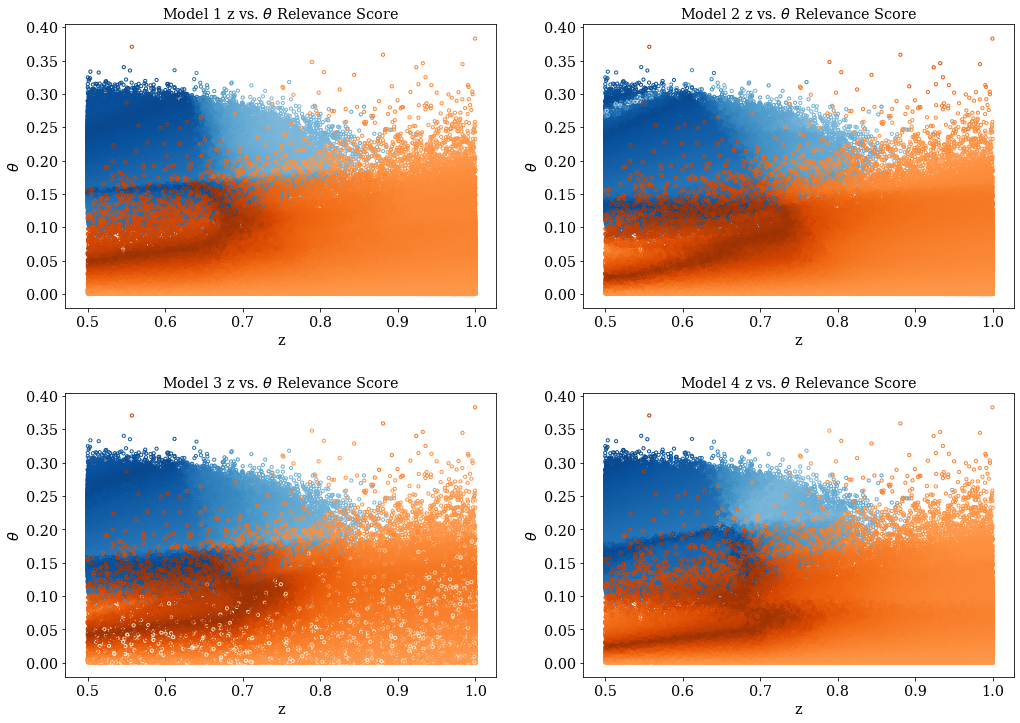}
    \caption{Relevance (LRP) scores for Toy Model 1D CNN for four separately trained models. The momentum fraction $z$ is shown along the horizontal axis while the angle $\theta$ is shown on the vertical axis. The orange distribution is the background while the blue distribution is the signal. The intensity of the color indicates the summed LRP score, with darker colors representing more confident predictions of the model.}
    \label{fig:Toy_XAUG_z_vs_theta_rel-z_model1}
\end{figure}

We plot the predicted score for these 4 models for $\theta$ versus $z$ in Fig.~\ref{fig:Toy_XAUG_z_vs_theta_predict_model1}. 
The orange distribution is the "background" while the blue distribution is the "signal". The intensity of the color indicates the predicted score. The plot in Fig.~\ref{fig:Toy_XAUG_z_vs_theta_rel-z_model1} shows the same events, but this time shading the histograms by the $z$ LRP score. 
In Fig.~\ref{fig:Toy_XAUG_z_vs_theta_predict_model1}, a clear gradient is shown across the decision boundary between the two populations. This is very consistent across different trainings of the network. Figure~\ref{fig:Toy_XAUG_z_vs_theta_rel-z_model1} shows high LRP scores along certain subspaces that correspond to details of decision boundaries, and while the various trainings are qualitatively similar in their "broad" features, the individual details vary among them. 
This indicates that there are subspaces within the classification that are not identical between different trainings. 

\begin{figure}[!ht]
    \begin{subfigure}{.5\textwidth}
        \centering
        \includegraphics[width=0.99\textwidth]{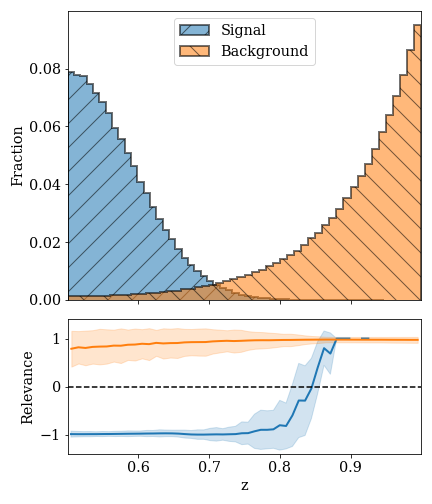}
        \caption{$z$ profile}
        \label{fig:2DLRP_profilez}
    \end{subfigure}
    \begin{subfigure}{.5\textwidth}
        \centering
        \includegraphics[width=0.99\textwidth]{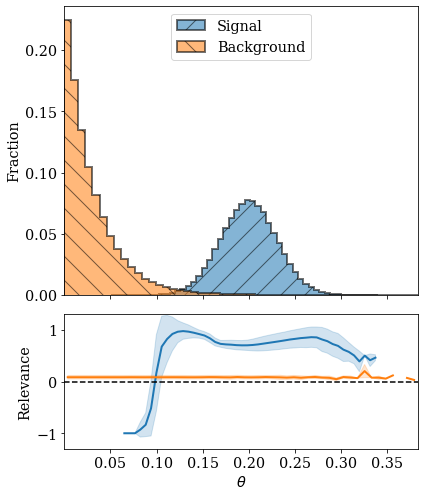}
        \caption{$\theta$ profile}
        \label{fig:2DLRP_profile_theta}
    \end{subfigure}
    \caption{Profiles for Toy 2D CNN Model with relevance and predictions averaged over the four models previously shown. We notice that z($\theta$) has the highest relevance to classify the event as background (signal), however the combination of the two variables (Fig. \ref{fig:Toy_XAUG_z_vs_theta_predict_model1}-\ref{fig:Toy_XAUG_z_vs_theta_rel-z_model1}) helps establish a clear decision boundary between both cases.}
    \label{fig:toy_profile_z-theta}
\end{figure}

Another way to display this information is shown in Fig.~\ref{fig:toy_profile_z-theta}. Here, the XAUG variables $z$ and $\theta$ are shown as 1D histograms, with their LRP relevance shown in the panel below. The signal is shown in blue and the background is shown in orange. The network is placing a high magnitude of relevance on the values that are far from the overlap region ($z > 0.75$, and $\theta > 0.10$).

\begin{figure}[!ht]
    \centering
    \includegraphics[width=0.48\textwidth]{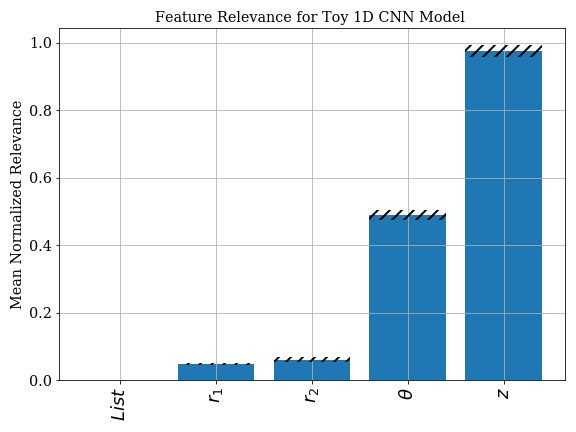}
    \includegraphics[width=0.48\textwidth]{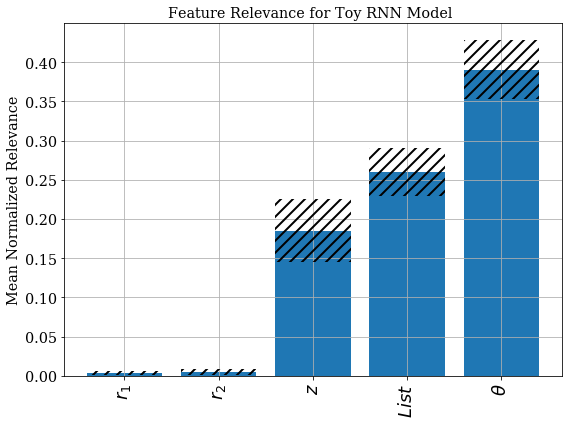}
    \caption{Mean normalized relevance score for 4 trainings of the toy 1D CNN (toy 1D RNN) on the left(right), where the relevance scores of constituent variables List ($\delta\alpha_1$, $\delta\beta_1$, $\delta\alpha_1$, $\delta\alpha_2$, $z_{const,1}$, $z_{const,2}$) are summed.}
    \label{fig:toybarplots}
\end{figure}
\begin{figure}[!ht]
    \centering
        \includegraphics[width=0.48\textwidth]{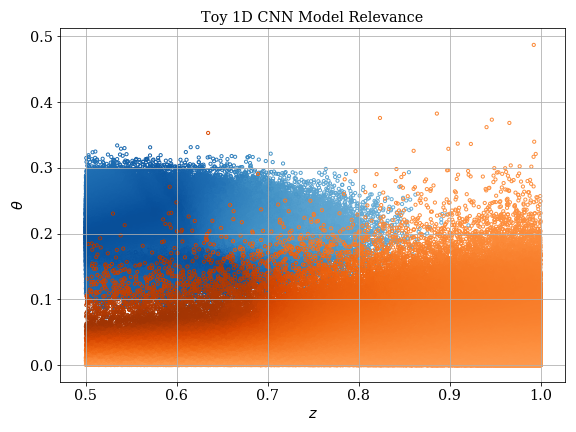}
        \includegraphics[width=0.48\textwidth]{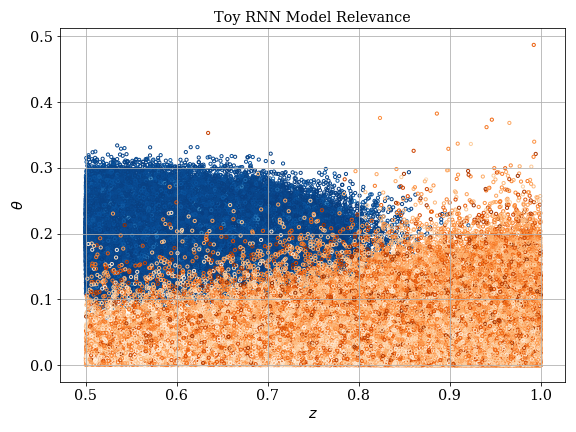}
    \caption{Mean LRP scores for z and $\theta$ variables in the Toy 1D CNN Model (left), and RNN (right).}
    \label{fig:toy1Dscatter}
\end{figure}

We performed the same studies shown above on the 1D CNN model. However, for brevity we show an average over 4 different trainings in our plots, with the RMS of the 4 trainings as the uncertainty. Similar to the 2D CNN model, we can see from Fig.~\ref{fig:toybarplots} that the variables given the most relevance by LRP were those with the greatest separation in combined phase space: $\z$ and $\theta$. We can see a clear decision boundary in their combined phase space in Fig.~\ref{fig:toy1Dscatter}. The left image again shows a clear gradient in the relevance of the variables, indicating that the network is unsure how to categorize the event in the overlap region.



The final model that we investigated was an RNN. This model showed much different relevance results from the other two, despite having comparable accuracy to the 1D and 2D CNN's. The RNN gives the greatest relevances to the constituent list variables, driven by the relevances of $z_1$ (the values of $\z$ broken up among the leading jet constituents) and $z_2$ (the values of $(1-\z)$ broken up among the subleading jet constituents), as can be seen in the right plot of Fig.~\ref{fig:toybarplots}. Although the CNN and RNN have different variables with greatest relevance, $z_1$ and $z_2$ encode the same information as $z$, just in constituent format. The RNN and CNN then still have the same leading variable, but the RNN gives more relevance to the variable in its constituent format rather than its event-level, or XAUG variable, format.

For the 2D CNN (Fig.~\ref{fig:Toy_XAUG_z_vs_theta_rel-z_model1}) and the 1D CNN (left image in Fig.~\ref{fig:toy1Dscatter}), the combined $\z-\theta$ phase space shows a clear decision boundary. The RNN, on the other hand, has a less apparent gradient (right image in Fig.~\ref{fig:toy1Dscatter}), with most of the relevance apparent in the background. The unclear decision boundary, in addition to the greater variation in the mean normalized relevance score, is most likely due to the implementation of the RNN's time axis, which currently couples all low-level features of two particles at a time. A possibly better alternative would be to couple one feature at a time for many particles; however, this would have been much more computationally intensive, and the RNN is already a factor of ten less computationally performant than its convolutional counterparts.

The figures in this section provide two important pieces of information. Firstly, in the case where there are XAUG variables that dominantly capture the information in the system, the LRP accurately selects the most important variables, rendering the image redundant, as seen in Fig.~\ref{fig:Toy_XAUG_featureSignificance_multimodel}. Secondly, the network can arrive at different optimizations, each with different subleading relevance, which can impact the confidence of the prediction in regions of confusion. However, these separately trained models have nearly identical performance (see Fig.~\ref{fig:Toy_XAUG_z_vs_theta_predict_model1}), despite clearly arriving at different minima (see Fig.~\ref{fig:Toy_XAUG_z_vs_theta_rel-z_model1}). 

These are not related to insufficient training data, as we have checked this with both low-and high-statistics samples and the features persist. This leads to a recommendation about numerical uncertainty in the classifier. 

\subsection{Numerical uncertainty in the classifier}

In the above studies, the LRP distributions of the various XAUG variables show that the main decision boundaries made by the classifier are relatively stable. However, there are details in the local approximation (LRP) boundaries that are caused by differences in relevance among subleading features. 

The LRP score is a reliable way to quantify the most relevant variables. However, this also means that the optimization with respect to less relevant variables is not easily determined, with very different local behavior for different trainings. This points to the need for a numerical uncertainty of classifiers that is not easily visible in the final classifier output, since the latter is primarily sensitive to the most relevant variables. Decisions based on less relevant variables can vary depending on the training. Optimizers like ADAM~\cite{adam} have a difficult time optimizing~\cite{adam_convergence} along these axes because they are so shallow, and the optimizer loss functions are dominated by the most relevant variables. 

A consequence of this study is that it is very important to perform multiple different trainings for networks, especially if the decisions are made on subdominant features. The trustworthiness of individual classifier decisions is very high in the case of differences in features with high relevance, but is considerably lower for features with low relevance. 

In the remaining study, we will therefore train the network 4 times and take the average response. The RMS of the predictions can then be used also for an uncertainty quantification, which we add for the relevance scores. These have similar interpretations to other systematic uncertainties, inasmuch as differences that exist, but are covered within the uncertainty, are not trustworthy differences.


\subsection{Particle Model Explanations}

The Toy Model introduced previously is intended as a highly simplified "cartoon" of jet substructure. The angle $\theta$ and momentum fraction $z$ of the "subjets" in the Toy Model are analogous to the $\Delta R$ and momentum fraction $z$ of the actual subjets from the soft drop algorithm. The drastic separation observed in the Toy Model will be reduced, although we can use the insight and techniques developed to apply to the particle-level simulation. As seen in the previous section, the toy model information can be completely exhausted by XAUG variables. However, for the more realistic particle model, as will be shown below, the XAUG variables do not fully exhaust the discrimination ability, but can be used to augment the performance of the low-level features. 

\begin{figure}[!ht]
    \centering
    \begin{subfigure}{0.32\textwidth}
        \includegraphics[width=\textwidth]{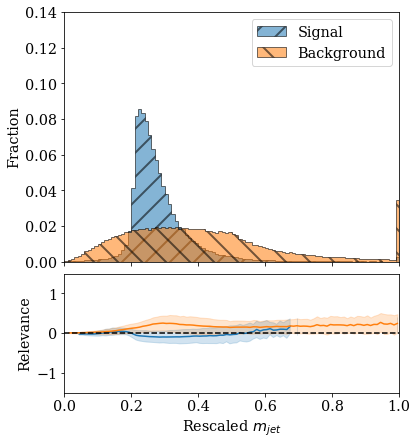}
    \end{subfigure}
    \begin{subfigure}{0.32\textwidth}
        \includegraphics[width=\textwidth]{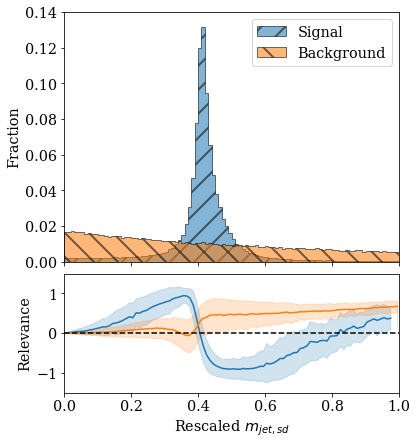}
    \end{subfigure} \\
    \begin{subfigure}{0.32\textwidth}
        \includegraphics[width=\textwidth]{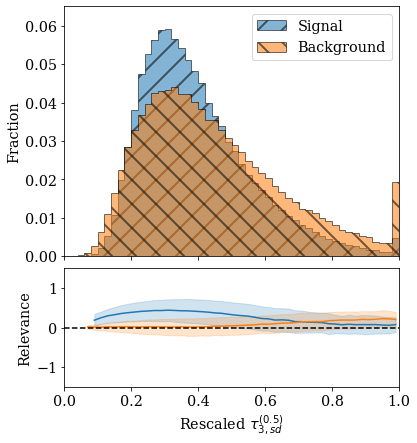}
    \end{subfigure}
     \begin{subfigure}{0.32\textwidth}
        \includegraphics[width=\textwidth]{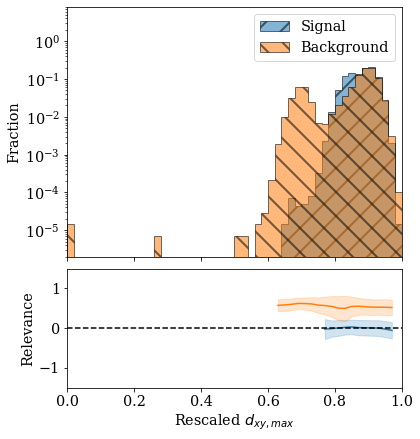}
    \end{subfigure} \\
    \caption{Histograms with profiles of LRP relevances normalized per event and averaged over four models. Signal and background are the $Z{\bbbar}$ and QCD samples, respectively.}
    \label{fig:2DCNN-topvariables}
\end{figure}

\begin{figure}[!ht]
 \centering
    \includegraphics[width=\linewidth]{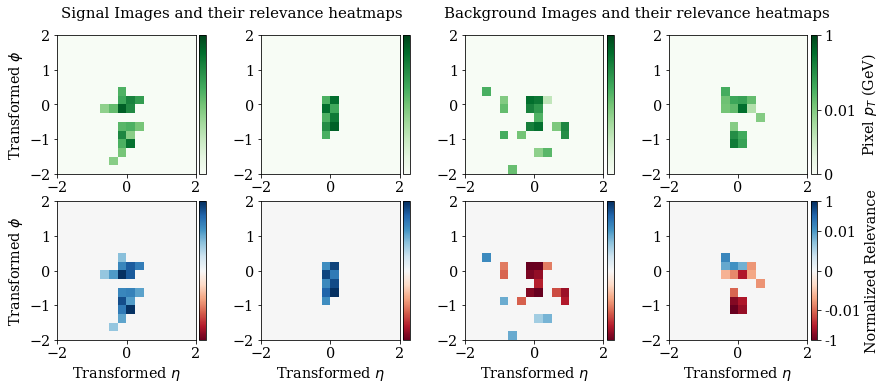}
    \caption{Individual Particle Model signal (left two columns) and background (right two columns) images (top) and their corresponding LRP score heatmaps (bottom).}
    \label{fig:pythiaimageslrp_individual}
\end{figure}

Distributions of the normalized variables are shown in the upper portions of some variables in Fig.~\ref{fig:2DCNN-topvariables} and in the Appendix for all variables in~\ref{fig:2DCNN-LRP1}-\ref{fig:2DCNN-LRP6}.
The jet images from a few individual events, along with their LRP heatmaps, are shown in Fig.~\ref{fig:pythiaimageslrp_individual}. The background events are more spread out than the signal events, which tend to be clustered among one or two subjets. This is exploited by the network, as can be observed in the bottom panes. The network identifies signal by weighting toward the central two subjets, while the network identifies background using pixels away from the two subjets. These images provide similar information to those in Ref.~\cite{deOliveira:2015xxd}, which show the Pearson Correlation Coefficient. 

\begin{figure}[!ht]
 \centering

        \includegraphics[width=0.45\linewidth]{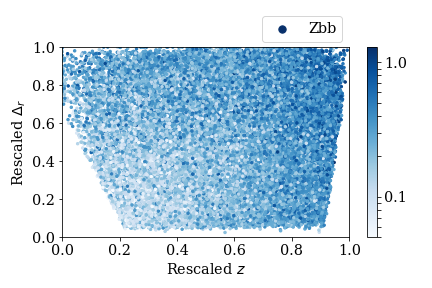}
        \includegraphics[width=0.45\linewidth]{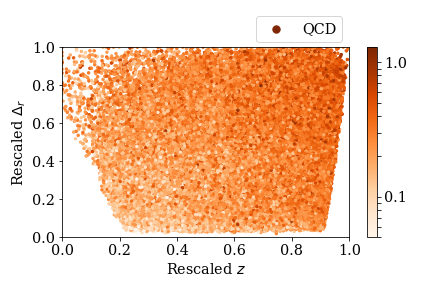}

    \caption{Averaged relevance score of 5 models for the particle list CNN for  $\Delta R_{subjets}$ vs $z$ for the $Z\bbbar$ and QCD samples. Analogous to the gradient plots of $\theta$ vs $z$ in Fig.~\ref{fig:Toy_XAUG_z_vs_theta_predict_model1} for the Toy 1DCNN.}
    \label{fig:1DCNN_scatter_dr_z}
\end{figure}

The predicted scores and LRP scores for the 2D CNN trained on the Particle Model are shown in Fig.~\ref{fig:1DCNN_scatter_dr_z} for the rescaled angular separation $\Delta R_{subjets}$ versus the rescaled momentum fraction of the subjets $z$. These are the analogous plots to our cartoon toy model, shown in Fig.~\ref{fig:1DCNN_scatter_dr_z}. It is clear that the separation between signal and background is reduced and the separation in these geometric variables is not as clear to the eye. 

\begin{figure}[!ht]
 \centering
       
        \includegraphics[width=0.45\linewidth]{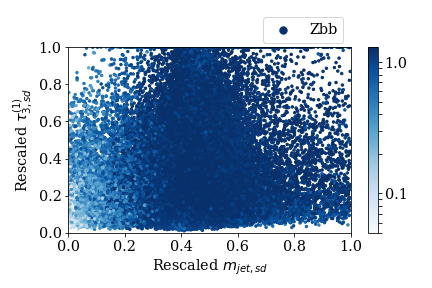}
        \includegraphics[width=0.45\linewidth]{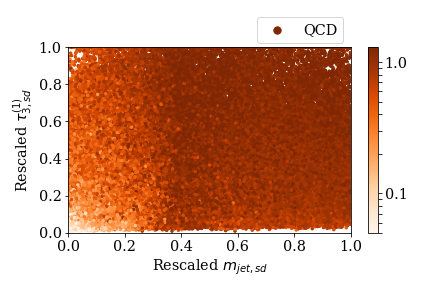}

    \caption{Averaged relevance score of 5 models for the particle list CNN for $\tau_{3, SD}^{1.0}$ vs $m_{jet, SD}$ for the $Z\bbbar$ and QCD samples. The features to compare are selected from Figure \ref{fig:1Dcnn_xaug_compare}.}
    \label{fig:1DCNN_scatter_msd_tau1}
\end{figure}

\begin{figure}[!ht]
 \centering
       
        \includegraphics[width=0.45\linewidth]{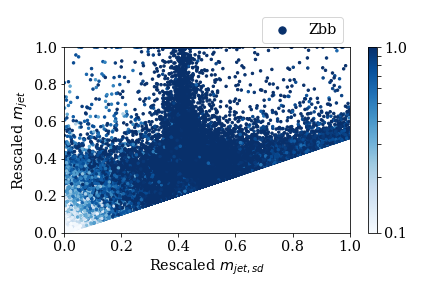}
        \includegraphics[width=0.45\linewidth]{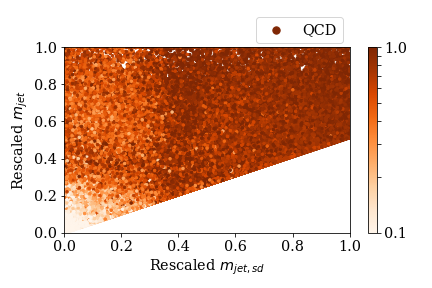}

    \caption{Averaged relevance score over 5 models for particle list CNN for ungroomed mass $m_{jet}$ vs groomed mass $m_{jet, SD}$ for the $Z\bbbar$ and QCD samples.}
    \label{fig:1DCNN_scatter_msd_mass}
\end{figure}

The simple intuition from viewing XAUG variable distributions can be translated into a choice of more appropriate variables. In particular, Fig.~\ref{fig:1DCNN_scatter_msd_tau1} shows the analogous distribution for two XAUG variables, the rescaled groomed jet mass $m_{SD}$ and the rescaled $\tau_{3,SD}^{1}$. It is clear that there are separations between the signal and background populations. The region for the rescaled $m_{SD}$ between 0.35 and 0.4 (corresponding to the Z mass window) tends to be preferred as a signal region, and just outside that region is preferred as background. In addition, there is a clear correlation between $m_{SD}$ and $\tau_{3,SD}^{1}$ that the network utilizes to identify background in the upper left and lower right portions of the plot (low $\tau$ at high mass, or high $\tau$ at low mass). Furthermore, Fig.~\ref{fig:1DCNN_scatter_msd_mass} shows the groomed versus ungroomed jet mass. This gives information about the fraction of the jet that is groomed away compared to the original, which can disentangle the hard and soft parts of the jet. The network clearly makes use of this information. In both of these plots, the network has regions of extremely high confidence that offer clear distinction between signal and background. 

Another way to see the same information is to look at the individual variables from the 2D CNN trained on the Particle Model in Fig.~\ref{fig:2DCNN-topvariables} for highly relevant variables, and in Figs.~\ref{fig:2DCNN-LRP1}-~\ref{fig:2DCNN-LRP6} in the appendix for all variables. 
These plots show the distribution of the XAUG variables in the top pane, as well as a profile distribution of their relevance scores in the bottom pane. The relevance score shows the contribution to the correct classification of either signal or background.

\begin{figure}[!ht]
    \centering
    \includegraphics[width=0.9\textwidth]{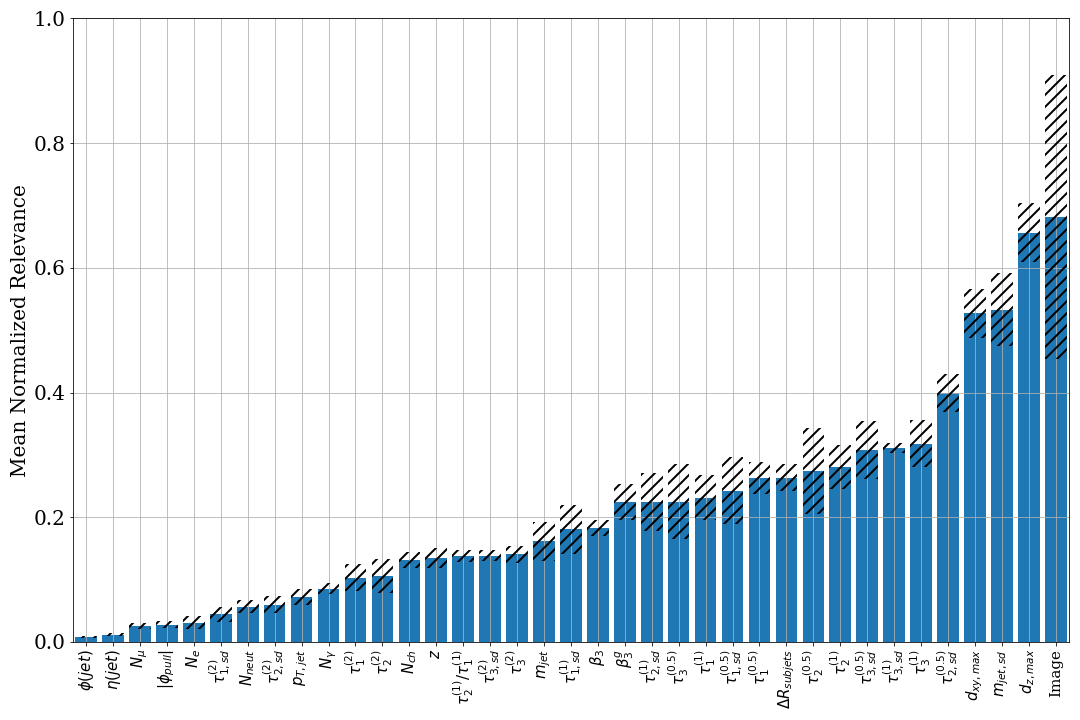}
    \caption{Input features with the greatest mean normalized relevance after averaging the relevance scores of four models for the 2D CNN. These models use $Z\bbbar$ and QCD simulation as signal and background, respectively. The relevance of the image is the summed relevance of all the pixels in the image.}
    \label{fig:xaug_bar_2DCNN}
\end{figure}

\begin{figure}[!ht]
    \centering
    \includegraphics[width=0.9\textwidth]{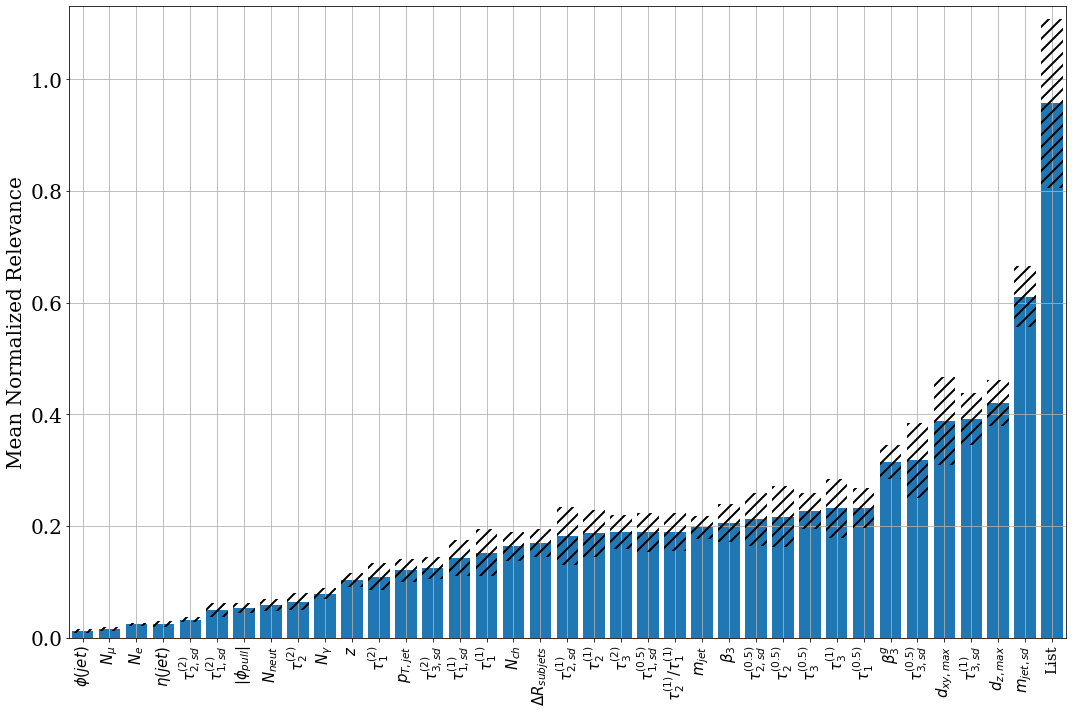}
    \caption{Input features with the greatest mean normalized relevance  after averaging the relevance scores of four models for the particle list CNN. "List" is the summed relevance of the 20 particle inputs per event and the 17 particle list inputs.}
    \label{fig:1Dcnn_xaug_compare}
\end{figure}

It is possible to use the information gathered by the LRP procedure to quantify the impact of each individual variable on the decision of the network. Figures~\ref{fig:xaug_bar_2DCNN} and~\ref{fig:1Dcnn_xaug_compare} show the mean relevance score over four models for each input in the 2D CNN and 1D CNN, respectively. In these cases, the relevance scores from the image and list are summed and displayed as a single bar in the histogram. The uncertainties on the scores are drawn from the RMS of four separate training iterations of the CNN. It is clear that there is a hierarchy of variables, with 5-6 very important ones with relevances above 0.3, around a dozen with moderate importance between 0.1 to 0.3, and around 5-6 with low relevances below 0.1. 

\begin{figure}[!ht]
    \centering
    \includegraphics[width=0.9\textwidth]{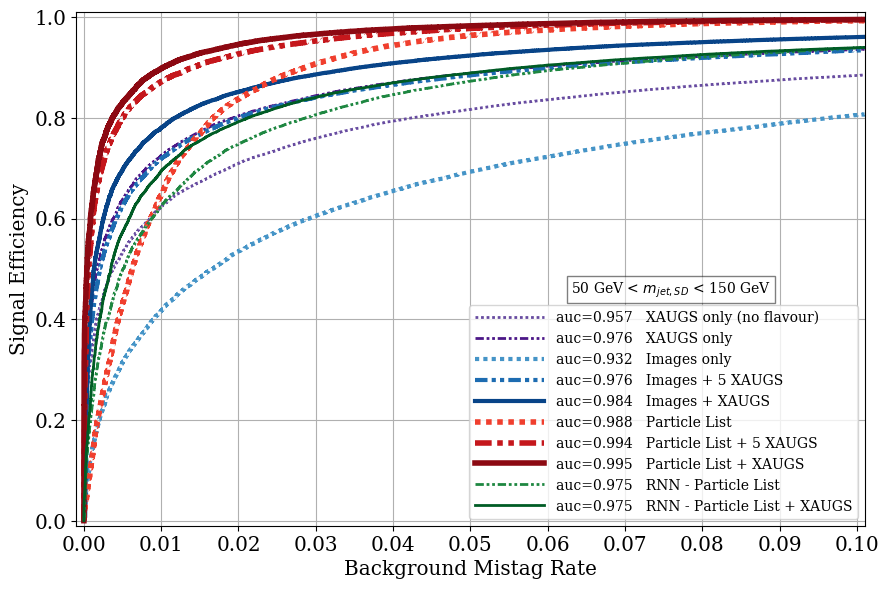}
    \caption{ROC curves from Image CNN, Particle List CNN, DNN with XAUG variables only, and RNN, with mass window cut of $50  < m_{jet, SD} < 150$ GeV.}
    \label{fig:roc_compare_all}
\end{figure}

In order to understand how much information is being used for the decision making, we can investigate the performance of our networks in several cases, shown in Fig.~\ref{fig:roc_compare_all}. Due to the fact that we are using a highly simplified model at the particle level only (for demonstration), the ROC curves are (over)optimistic about performance, although it is still instructive to investigate differences to understand relevant features.  

First, as a baseline, we can look at the simplest case of MLPs using only the XAUG variables as inputs. We will look at two such cases, one where we do not use flavor information (particle content, $d_{xy}$, and $d_{z}$), and then when this information is added. These are labeled "XAUGs only (no flavor)" and "XAUGs only"). 

Next, we can investigate the 2D CNN with only the jet images as input (labeled as "Images only" in the figure). We then combine the information together from the image and the 5 XAUG variables with highest relevance ("Images + 5 XAUGs") or with all XAUG variables ("Images + XAUGs").  

We can repeat the above cases with the particle list for the 1D CNN ("Particle List"), and then add the top 5 XAUG variables ("Particle List + 5 XAUGs") and all XAUG variables ("Particle List + XAUGs"). The RNN can also be similarly investigated with and without XAUG variables, shown in "RNN - Particle List" and "RNN - Particle List + XAUGs". 

We see that the image-only NN performs the worst, with an AUC of 0.932. The XAUG variables perform better, achieving AUC of 0.957 (0.976) without (with) flavor information. Combining the image with the XAUG variables achieves an AUC of 0.984. Combining the particle list with the XAUG variables achieves AUCs of 0.994 and 0.995 for the top 5, and all XAUG variables, respectively. Similarly, the RNN achieves an AUC=0.975 regardless of whether the XAUG variables are used. 

These results demonstrate that adding XAUG variables can considerably improve network performance. In fact, the XAUG-only performance is already fairly good at distinguishing between $Z\rightarrow \bbbar$ and QCD. It may be quite useful for some simplistic cases to consider these simple networks, guided by what DNNs find important. In cases where more discrimination is necessary, combining XAUG variables with the network can act as a sort of guide for the optimization, achieving better performance together than either achieves individually. It also gives a robust method for investigation of subspaces of relevance that can be understood easily by analyzers. 
Furthermore, the cases where we add only the 5 XAUG variables with the highest relevance scores are almost identical to using all XAUG variables. We can therefore conclude that the XAUG variables with the highest relevance scores encompass the primary decision factors that the network is making, and could be used by analysts to audit and understand network behavior. In addition, only these few XAUG variables would be needed to greatly improve performance. 

As can be seen in Fig.~\ref{fig:xaug_bar_2DCNN}, the LRP relevance of the image is greater than that of any one XAUG variable. This is a noticeable change from the toy model results in Fig.~\ref{fig:Toy_XAUG_featureSignificance_multimodel}, where the image ranks lower in relevance than the top XAUG variables. There, the XAUG variables capture all of the information in the event, and a single XAUG variable has greater relevance than the image alone. For the particle level model, the total information of the event can be captured by a subset of the XAUG inputs, since a subset's combined relevance score is greater than the image's score alone. The same conclusions can be drawn for the particle list CNN by comparing Fig.~\ref{fig:toybarplots} and Fig.~\ref{fig:1Dcnn_xaug_compare}.

In addition to the above observations, Fig.~\ref{fig:xaug_bar_2DCNN} shows considerable variation from one training to another in the local behavior of networks. The local network optimization depends weakly on features with low relevance, so decisions based on these features should be trusted to a somewhat lesser degree. These findings suggest that networks should be trained multiple times, with the average taken and variations counted as uncertainties. 
In our case, the networks tend to be strongly affected by dominant, smaller subspaces related to a few shape-based variables, jet mass with and without grooming, and the lifetime information. Such an analysis combining LRP and XAUG variables could also be applied in other use cases to rank salient features.

 %


\section{Conclusion}

\label{sec:conclusions}

We have presented a scheme to systematically explain jet identification decisions from a deep neural network (DNN) using eXpert AUGmented variables (XAUG) and layerwise relevance propagation (LRP). The combination of these two techniques can shed light on network decisions that would otherwise be difficult to ascertain. The XAUG variables can also be used alone for classification, or can be used in conjunction with lower-level / higher-dimensional classifiers to "guide" the network decisions, resulting in better performance. In some cases, such as with the toy model presented, XAUG variables can even capture the information of the lower-level networks entirely if there are subspaces of the network optimization that are correlated with the XAUG variables themselves. In other cases, such as the more realistic particle model presented, the low-level information sometimes provide additional discrimination power over the XAUG variables alone. Both have comparable performance in the studies presented. However, the best performance was observed when the two approaches were combined. 
The ranked relevance scores of inputs can give insight into which XAUG variables are best to replace or accompany the original input.

An additional benefit to using LRP and XAUG variables is the ability to investigate relevant subspaces of the training. This can highlight when there are very shallow optimizations, where training multiple times could lead to slightly different subspaces. In such cases, it is recommended to train multiple times and take the mean of the predictions. This provides a framework to compute numerical systematic uncertainties in the training.


\clearpage

\acknowledgments

This work was supported under NSF Grants PHY-1806573, PHY-1719690 and PHY-1652066. Computations were performed at the Center for Computational Research at the University at Buffalo.

We would like to acknowledge Kyle Cranmer, Marat Freytsis, Loukas Gouskos, Gregor Kasieczka, Martin Kwok, Simone Marzani, David Miller, Ben Nachman, Juska Pekkanen, Jesse Thaler, Daniel Whiteson, and David Yu for helpful conversations during the BOOST 2020 poster session and manuscript review. 

\appendix

\section{Software Versions}

We used {\tt PYTHIA} v8.2.35~\cite{Sjostrand:2014zea} with the default parameters, with events stored in the {\tt ROOT}~\cite{rene_brun_2019_3895860} format. Both the toy model and particle-level simulations were processed with {\tt coffea} v0.6~\cite{lindsey_gray_2020_4247940} using {\tt uproot} v3.11~\cite{jim_pivarski_2020_3952728} and {\tt AwkwardArray} v0.12.0rc1~\cite{jim_pivarski_2019_3275017}. We used {\tt tensorflow} v1.13~\cite{tensorflow2015-whitepaper} and {\tt iNNvestigate}~\cite{alber2018innvestigate} for the DNNs and LRP, respectively. 

Our software can be found \href{https://github.com/ubcms-xai/XAICoffea}{here} on github. 

\clearpage
\section{Relevance Scores for All XAUG Variables}

\begin{figure}[!ht]
    \centering
    \begin{subfigure}{0.32\textwidth}
        \includegraphics[width=\textwidth]{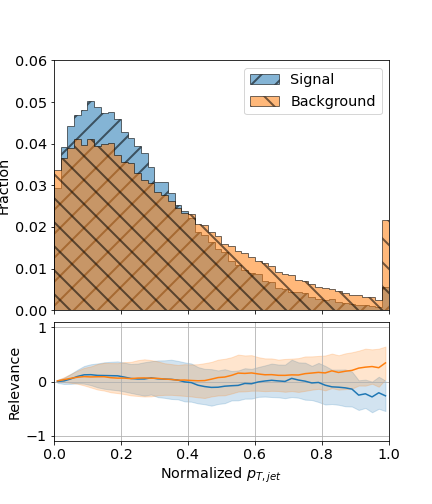}
    \end{subfigure}
    \begin{subfigure}{0.32\textwidth}
        \includegraphics[width=\textwidth]{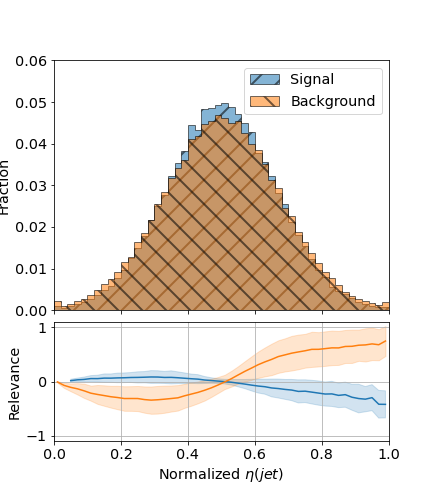}
    \end{subfigure}
    \begin{subfigure}{0.32\textwidth}
        \includegraphics[width=\textwidth]{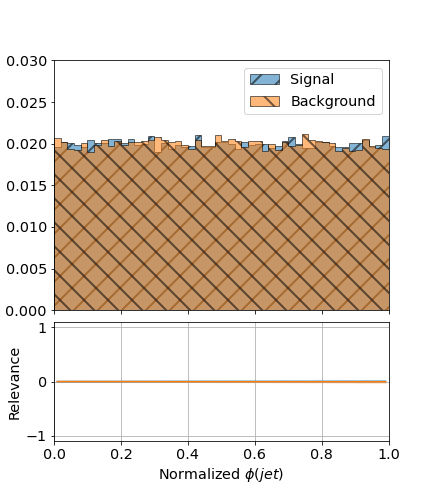}
    \end{subfigure} \\
    \begin{subfigure}{0.32\textwidth}
        \includegraphics[width=\textwidth]{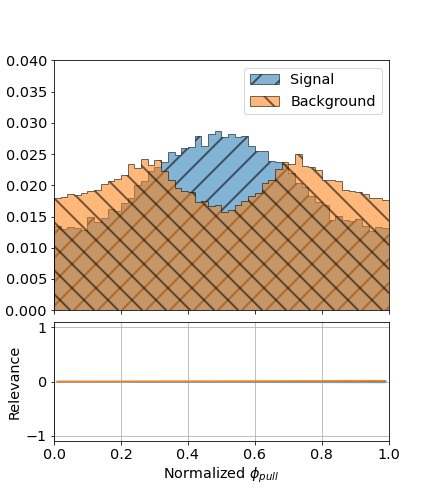}
    \end{subfigure} \\
    \caption{Histograms with profiles of normalized LRP relevances.}
    \label{fig:2DCNN-LRP1}
\end{figure}

\begin{figure}[!ht]
    \centering
    \begin{subfigure}{0.32\textwidth}
        \includegraphics[width=\textwidth]{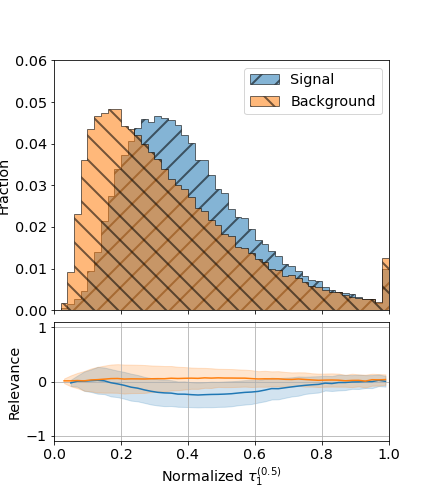}
    \end{subfigure}
    \begin{subfigure}{0.32\textwidth}
        \includegraphics[width=\textwidth]{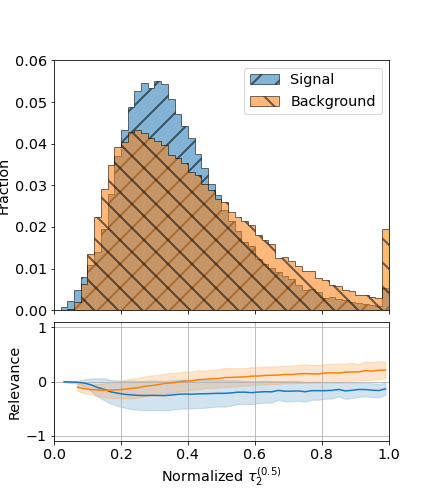}
    \end{subfigure}
    \begin{subfigure}{0.32\textwidth}
        \includegraphics[width=\textwidth]{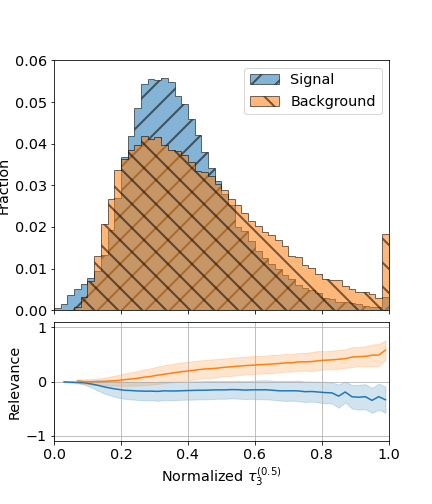}
    \end{subfigure} \\
        \begin{subfigure}{0.32\textwidth}
        \includegraphics[width=\textwidth]{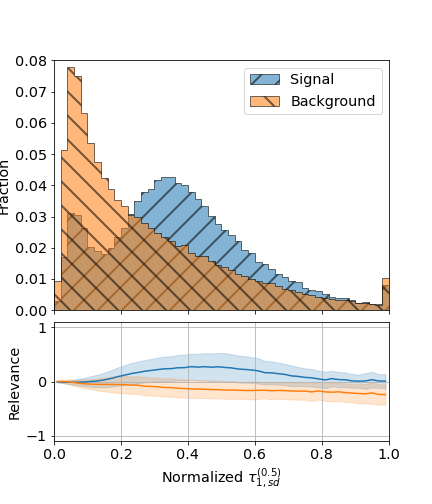}
    \end{subfigure}
    \begin{subfigure}{0.32\textwidth}
        \includegraphics[width=\textwidth]{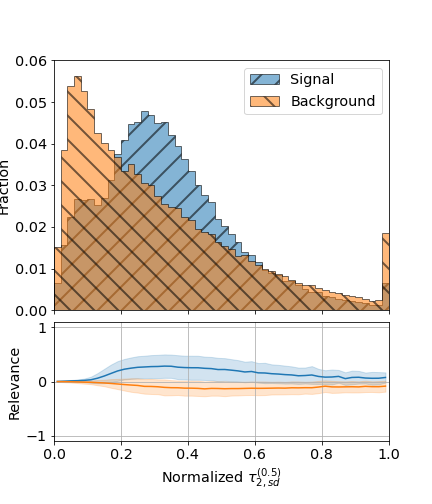}
    \end{subfigure}
    \begin{subfigure}{0.32\textwidth}
        \includegraphics[width=\textwidth]{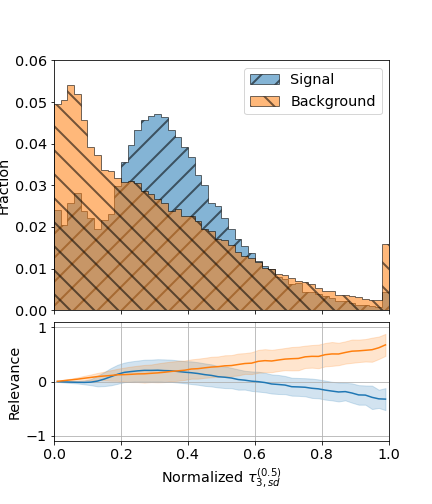}
    \end{subfigure} \\
    \caption{Histograms with profiles of normalized LRP relevances.}
    \label{fig:2DCNN-LRP2}
\end{figure}

\begin{figure}[!htb]
    \centering
    \begin{subfigure}{0.32\textwidth}
        \includegraphics[width=\textwidth]{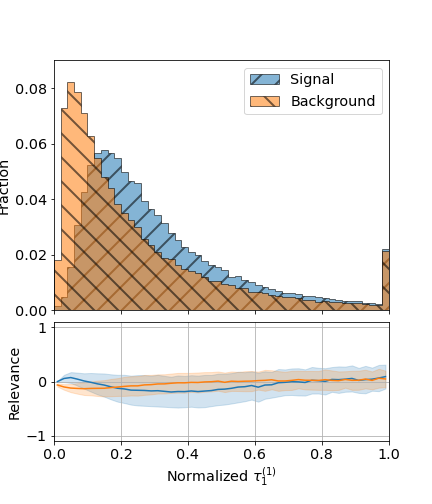}
    \end{subfigure}
    \begin{subfigure}{0.32\textwidth}
        \includegraphics[width=\textwidth]{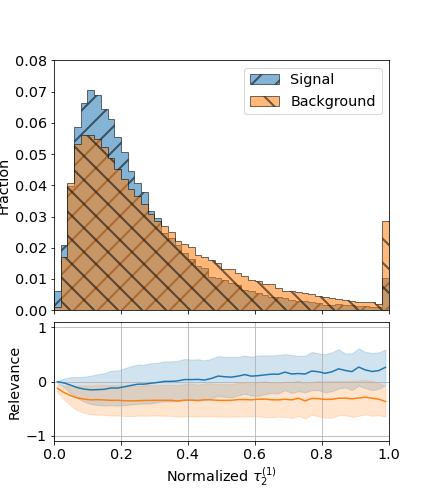}
    \end{subfigure}
    \begin{subfigure}{0.32\textwidth}
        \includegraphics[width=\textwidth]{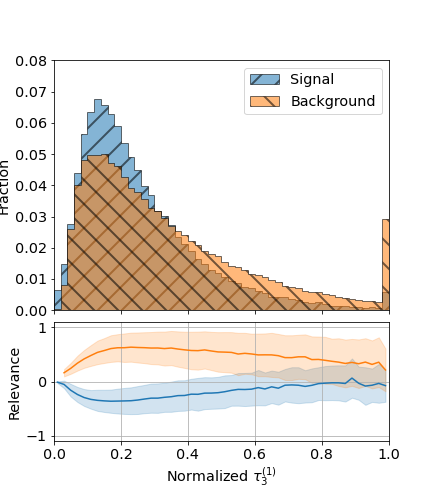}
    \end{subfigure} \\
        \begin{subfigure}{0.32\textwidth}
        \includegraphics[width=\textwidth]{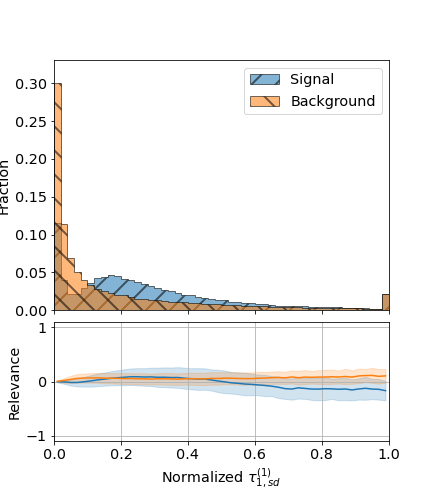}
    \end{subfigure}
    \begin{subfigure}{0.32\textwidth}
        \includegraphics[width=\textwidth]{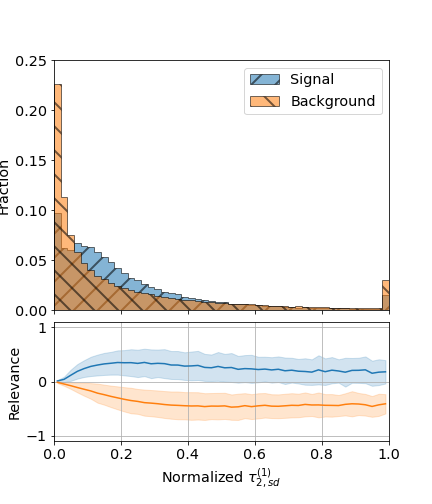}
    \end{subfigure}
    \begin{subfigure}{0.32\textwidth}
        \includegraphics[width=\textwidth]{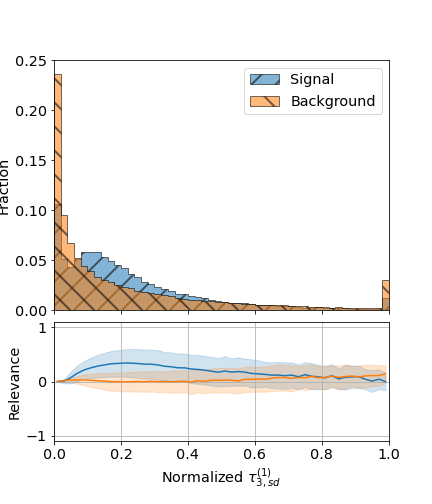}
    \end{subfigure} \\
    \caption{Histograms with profiles of normalized LRP relevances.}
    \label{fig:2DCNN-LRP3}
\end{figure}

\begin{figure}[!htb]
    \centering
    \begin{subfigure}{0.32\textwidth}
        \includegraphics[width=\textwidth]{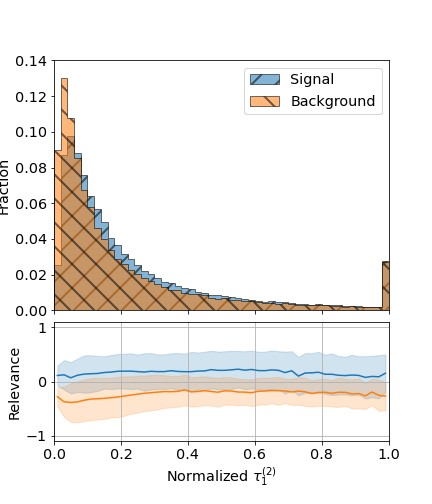}
    \end{subfigure}
    \begin{subfigure}{0.32\textwidth}
        \includegraphics[width=\textwidth]{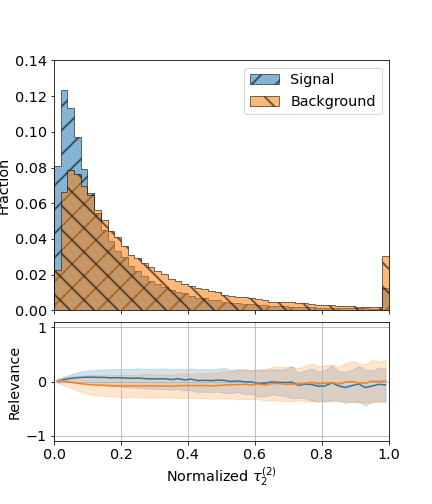}
    \end{subfigure}
    \begin{subfigure}{0.32\textwidth}
        \includegraphics[width=\textwidth]{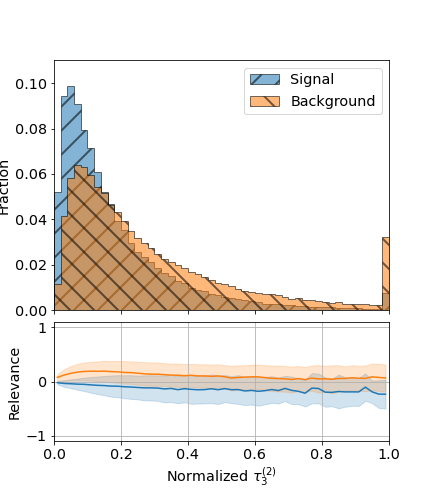}
    \end{subfigure} \\
        \begin{subfigure}{0.32\textwidth}
        \includegraphics[width=\textwidth]{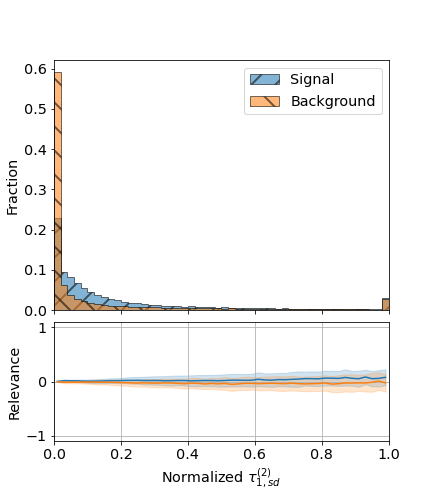}
    \end{subfigure}
    \begin{subfigure}{0.32\textwidth}
        \includegraphics[width=\textwidth]{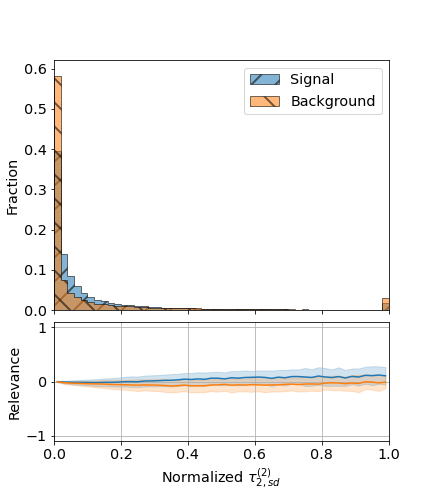}
    \end{subfigure}
    \begin{subfigure}{0.32\textwidth}
        \includegraphics[width=\textwidth]{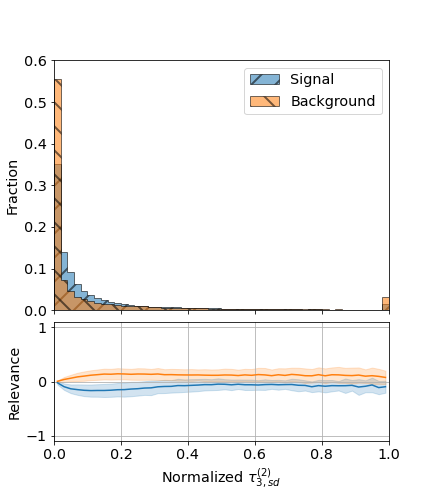}
    \end{subfigure} \\
    \caption{Histograms with profiles of normalized LRP relevances.}
    \label{fig:2DCNN-LRP4}
\end{figure}

\begin{figure}[!htb]
    \centering
    \begin{subfigure}{0.32\textwidth}
        \includegraphics[width=\textwidth]{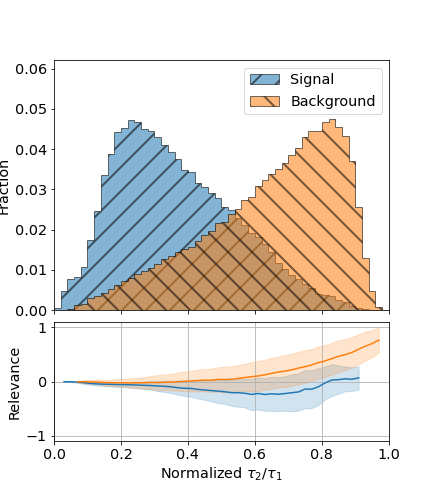}
    \end{subfigure}
    \begin{subfigure}{0.32\textwidth}
        \includegraphics[width=\textwidth]{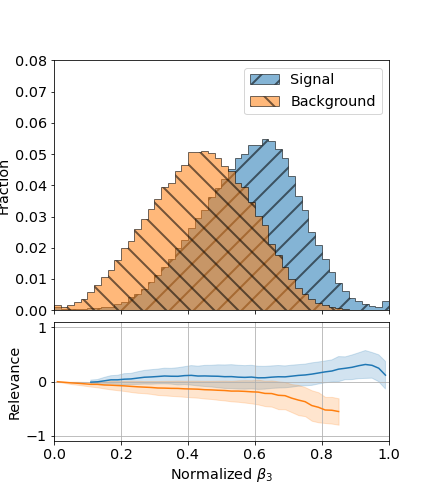}
    \end{subfigure}
    \begin{subfigure}{0.32\textwidth}
        \includegraphics[width=\textwidth]{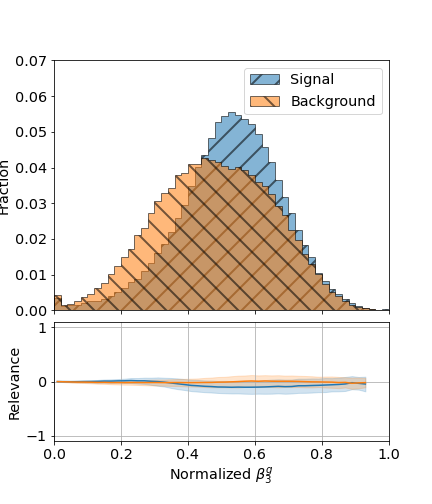}
    \end{subfigure}
    \caption{Histograms with profiles of normalized LRP relevances.}
    \label{fig:2DCNN-LRP5}
\end{figure}

\begin{figure}[!ht]
    \centering
    \begin{subfigure}{0.32\textwidth}
        \includegraphics[width=\textwidth]{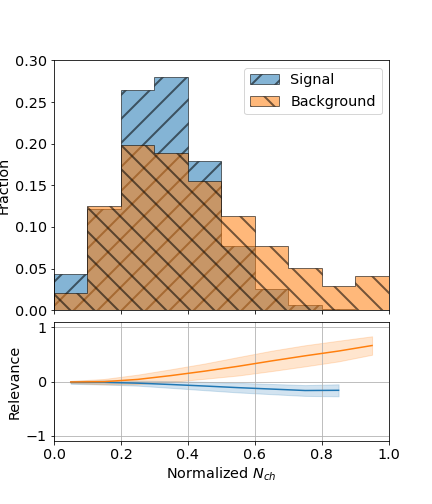}
    \end{subfigure}
    \begin{subfigure}{0.32\textwidth}
        \includegraphics[width=\textwidth]{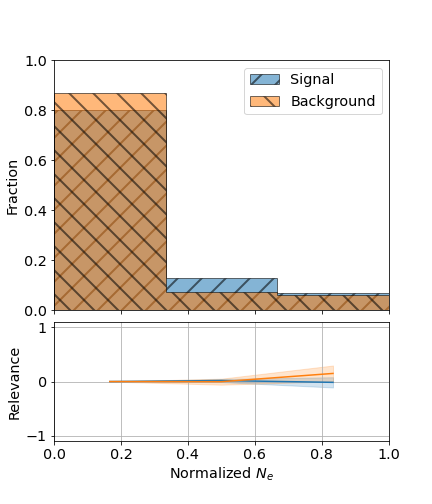}
    \end{subfigure}
    \begin{subfigure}{0.32\textwidth}
        \includegraphics[width=\textwidth]{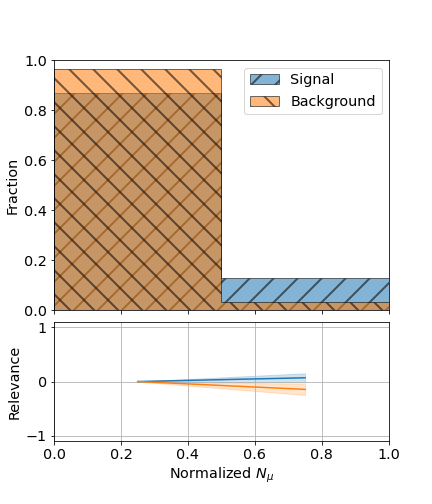}
    \end{subfigure} \\
        \begin{subfigure}{0.32\textwidth}
        \includegraphics[width=\textwidth]{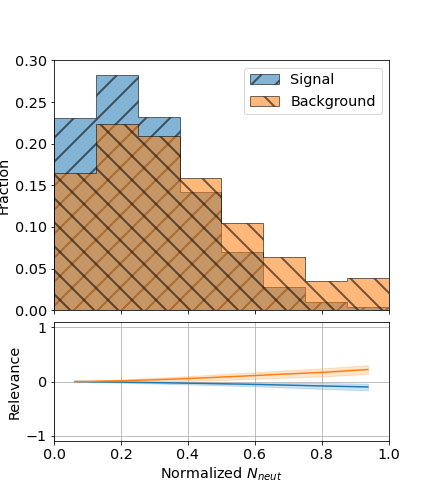}
    \end{subfigure}
    \begin{subfigure}{0.32\textwidth}
        \includegraphics[width=\textwidth]{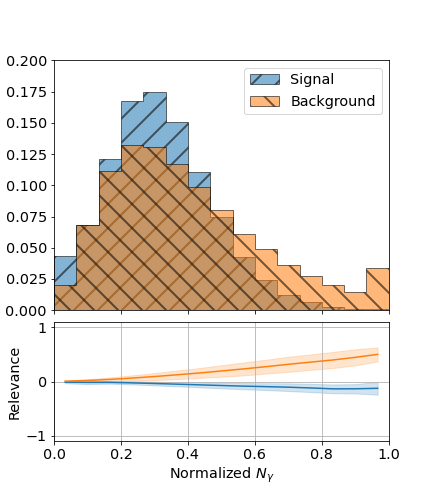}
    \end{subfigure} \\
    \caption{Histograms with profiles of normalized LRP relevances.}
    \label{fig:2DCNN-LRP6}
\end{figure}

\clearpage


\bibliographystyle{JHEP}
\bibliography{biblio}

\providecommand{\href}[2]{#2}\begingroup\raggedright\begin{thebibliography}{10}

\bibitem{Abdesselam:2010pt}
A.~Abdesselam et~al., {\it {Boosted Objects: A Probe of Beyond the Standard
  Model Physics}},  {\em Eur. Phys. J.} {\bf C71} (2011) 1661,
  [\href{https://arxiv.org/abs/1012.5412}{{\tt arXiv:1012.5412}}].

\bibitem{Altheimer:2012mn}
A.~Altheimer et~al., {\it {Jet Substructure at the Tevatron and LHC: New
  results, new tools, new benchmarks}},  {\em J. Phys.} {\bf G39} (2012)
  063001, [\href{https://arxiv.org/abs/1201.0008}{{\tt arXiv:1201.0008}}].

\bibitem{Altheimer:2013yza}
A.~Altheimer et~al., {\it {Boosted Objects and Jet Substructure at the LHC.
  Report of BOOST2012, held at IFIC Valencia, 23rd-27th of July 2012}},  {\em
  Eur. Phys. J.} {\bf C74} (2014), no.~3 2792,
  [\href{https://arxiv.org/abs/1311.2708}{{\tt arXiv:1311.2708}}].

\bibitem{Adams:2015hiv}
D.~Adams et~al., {\it {{Towards an Understanding of the Correlations in Jet
  Substructure}}},  {\em Eur. Phys. J.} {\bf C75} (2015), no.~9 409,
  [\href{https://arxiv.org/abs/1504.00679}{{\tt arXiv:1504.00679}}].

\bibitem{Larkoski:2017jix}
A.~J. Larkoski, I.~Moult, and B.~Nachman, {\it {Jet Substructure at the Large
  Hadron Collider: A Review of Recent Advances in Theory and Machine
  Learning}},  {\em Phys. Rept.} {\bf 841} (2020) 1--63,
  [\href{https://arxiv.org/abs/1709.04464}{{\tt arXiv:1709.04464}}].

\bibitem{Asquith:2018igt}
R.~Kogler et~al., {\it {Jet Substructure at the Large Hadron Collider:
  Experimental Review}},  {\em Rev. Mod. Phys.} {\bf 91} (2019), no.~4 045003,
  [\href{https://arxiv.org/abs/1803.06991}{{\tt arXiv:1803.06991}}].

\bibitem{Kasieczka:2019dbj}
A.~Butter et~al., {\it {The Machine Learning Landscape of Top Taggers}},  {\em
  SciPost Phys.} {\bf 7} (2019) 014,
  [\href{https://arxiv.org/abs/1902.09914}{{\tt arXiv:1902.09914}}].

\bibitem{Guest:2016iqz}
D.~Guest, J.~Collado, P.~Baldi, S.-C. Hsu, G.~Urban, and D.~Whiteson, {\it {Jet
  Flavor Classification in High-Energy Physics with Deep Neural Networks}},
  {\em Phys. Rev.} {\bf D94} (2016), no.~11 112002,
  [\href{https://arxiv.org/abs/1607.08633}{{\tt arXiv:1607.08633}}].

\bibitem{Murdoch_2019}
W.~J. Murdoch, C.~Singh, K.~Kumbier, R.~Abbasi-Asl, and B.~Yu, {\it
  Definitions, methods, and applications in interpretable machine learning},
  {\em Proceedings of the National Academy of Sciences} {\bf 116} (Oct, 2019)
  22071–22080.

\bibitem{DBLP:journals/corr/Lipton16a}
Z.~C. Lipton, {\it {The Mythos of Model Interpretability: In Machine Learning,
  the Concept of Interpretability is Both Important and Slippery.}},  {\em
  Queue} {\bf 16} (June, 2018) 31–57.
  \url{https://doi.org/10.1145/3236386.3241340}.

\bibitem{Goodfellow-et-al-2016}
I.~Goodfellow, Y.~Bengio, and A.~Courville, {\em Deep Learning}.
\newblock MIT Press, 2016.
\newblock \url{http://www.deeplearningbook.org}.

\bibitem{DBLP:journals/corr/RibeiroSG16}
M.~T. Ribeiro, S.~Singh, and C.~Guestrin, {\it {"Why Should I Trust You?":
  Explaining the Predictions of Any Classifier}},  in {\em Proceedings of the
  22nd ACM SIGKDD International Conference on Knowledge Discovery and Data
  Mining}, KDD16, (New York, NY, USA), p.~1135–1144, Association for
  Computing Machinery, 2016.
\newblock \href{https://arxiv.org/abs/1602.04938}{{\tt arXiv:1602.04938}}.
\newblock \url{https://doi.org/10.1145/2939672.2939778}.

\bibitem{SLISE}
A.~Bj{\"o}rklund, A.~Henelius, E.~Oikarinen, K.~Kallonen, and K.~Puolam{\"a}ki,
  {\it {{Sparse Robust Regression for Explaining Classifiers}}},  in {\em
  Discovery Science} (P.~Kralj~Novak, T.~{\v{S}}muc, and S.~D{\v{z}}eroski,
  eds.), (Cham), pp.~351--366, Springer International Publishing, 2019.

\bibitem{10.1371/journal.pone.0130140}
S.~Bach, A.~Binder, G.~Montavon, F.~Klauschen, K.-R. MÃŒller, and W.~Samek,
  {\it {On Pixel-Wise Explanations for Non-Linear Classifier Decisions by
  Layer-Wise Relevance Propagation}},  {\em PLOS ONE} {\bf 10} (07, 2015)
  1--46.

\bibitem{DBLP:journals/corr/abs-1708-08296}
W.~Samek, T.~Wiegand, and K.-R. Müller, {\it {Explainable Artificial
  Intelligence: Understanding, Visualizing and Interpreting Deep Learning
  Models}},  {\em ITU Journal: ICT Discoveries - Special Issue 1 - The Impact
  of Artificial Intelligence (AI) on Communication Networks and Services} {\bf
  1} (10, 2017) 1--10, [\href{https://arxiv.org/abs/1708.08296}{{\tt
  arXiv:1708.08296}}].

\bibitem{Montavon2019}
G.~Montavon, A.~Binder, S.~Lapuschkin, W.~Samek, and K.-R. M{\"u}ller, {\em
  Layer-Wise Relevance Propagation: An Overview}, pp.~193--209.
\newblock Springer International Publishing, Cham, 2019.

\bibitem{cf736d955d6e4296a9b7255bfee3b403}
D.~Baehrens, T.~Schroeter, S.~Harmeling, M.~Kawanabe, K.~Hansen, and K.~Muller,
  {\it How to explain individual classification decisions},  {\em Journal of
  Machine Learning Research} {\bf 11} (6, 2010) 1803--1831.

\bibitem{Datta:2017rhs}
K.~Datta and A.~Larkoski, {\it {How Much Information is in a Jet?}},  {\em
  JHEP} {\bf 06} (2017) 073, [\href{https://arxiv.org/abs/1704.08249}{{\tt
  arXiv:1704.08249}}].

\bibitem{Lim:2018toa}
S.~H. Lim and M.~M. Nojiri, {\it {Spectral Analysis of Jet Substructure with
  Neural Networks: Boosted Higgs Case}},  {\em JHEP} {\bf 10} (2018) 181,
  [\href{https://arxiv.org/abs/1807.03312}{{\tt arXiv:1807.03312}}].

\bibitem{Chakraborty:2019imr}
A.~Chakraborty, S.~H. Lim, and M.~M. Nojiri, {\it {Interpretable deep learning
  for two-prong jet classification with jet spectra}},  {\em JHEP} {\bf 07}
  (2019) 135, [\href{https://arxiv.org/abs/1904.02092}{{\tt
  arXiv:1904.02092}}].

\bibitem{Chen:2019apv}
K.-F. Chen and Y.-T. Chien, {\it Deep learning jet substructure from
  two-particle correlations},  {\em Physical Review D} {\bf 101} (Jun, 2020).

\bibitem{Kasieczka:2020nyd}
G.~Kasieczka, S.~Marzani, G.~Soyez, and G.~Stagnitto, {\it Towards machine
  learning analytics for jet substructure},  {\em JHEP} {\bf 2020} (Sep, 2020).

\bibitem{Aad:2019aic}
{\bf ATLAS} Collaboration, G.~Aad et~al., {\it {ATLAS b-jet identification
  performance and efficiency measurement with $t{\bar{t}}$ events in pp
  collisions at $\sqrt{s}=13$ TeV}},  {\em Eur. Phys. J. C} {\bf 79} (2019),
  no.~11 970, [\href{https://arxiv.org/abs/1907.05120}{{\tt
  arXiv:1907.05120}}].

\bibitem{Aad:2015ydr}
{\bf ATLAS} Collaboration, G.~Aad et~al., {\it {Performance of $b$-Jet
  Identification in the ATLAS Experiment}},  {\em JINST} {\bf 11} (2016),
  no.~04 P04008, [\href{https://arxiv.org/abs/1512.01094}{{\tt
  arXiv:1512.01094}}].

\bibitem{Aaboud:2018xwy}
{\bf ATLAS} Collaboration, M.~Aaboud et~al., {\it {Measurements of b-jet
  tagging efficiency with the ATLAS detector using $ t\overline{t} $ events at
  $ \sqrt{s}=13 $ TeV}},  {\em JHEP} {\bf 08} (2018) 089,
  [\href{https://arxiv.org/abs/1805.01845}{{\tt arXiv:1805.01845}}].

\bibitem{Aad:2019uoz}
{\bf ATLAS} Collaboration, G.~Aad et~al., {\it {Identification of boosted Higgs
  bosons decaying into $b$-quark pairs with the ATLAS detector at 13 $\text
  {TeV}$}},  {\em Eur. Phys. J. C} {\bf 79} (2019), no.~10 836,
  [\href{https://arxiv.org/abs/1906.11005}{{\tt arXiv:1906.11005}}].

\bibitem{Sirunyan:2017ezt}
{\bf CMS} Collaboration, A.~Sirunyan et~al., {\it {Identification of
  heavy-flavour jets with the CMS detector in pp collisions at 13 TeV}},  {\em
  JINST} {\bf 13} (2018), no.~05 P05011,
  [\href{https://arxiv.org/abs/1712.07158}{{\tt arXiv:1712.07158}}].

\bibitem{CMS:2019gpd}
{\bf CMS} Collaboration, A.~M. Sirunyan et~al., {\it {Identification of heavy,
  energetic, hadronically decaying particles using machine-learning
  techniques}},  {\em JINST} {\bf 15} (2020), no.~06 P06005,
  [\href{https://arxiv.org/abs/2004.08262}{{\tt arXiv:2004.08262}}].

\bibitem{Salam:2009jx}
G.~P. Salam, {\it {{Towards Jetography}}},  {\em Eur. Phys. J. C} {\bf 67}
  (2010) 637--686, [\href{https://arxiv.org/abs/0906.1833}{{\tt
  arXiv:0906.1833}}].

\bibitem{Kasieczka:2017nvn}
G.~Kasieczka, T.~Plehn, M.~Russell, and T.~Schell, {\it {Deep-learning Top
  Taggers or The End of QCD?}},  {\em JHEP} {\bf 05} (2017) 006,
  [\href{https://arxiv.org/abs/1701.08784}{{\tt arXiv:1701.08784}}].

\bibitem{Moreno:2019neq}
E.~A. Moreno, T.~Q. Nguyen, J.-R. Vlimant, O.~Cerri, H.~B. Newman, A.~Periwal,
  M.~Spiropulu, J.~M. Duarte, and M.~Pierini, {\it {Interaction networks for
  the identification of boosted $H \rightarrow b\overline{b}$ decays}},  {\em
  Phys. Rev. D} {\bf 102} (2020), no.~1 012010,
  [\href{https://arxiv.org/abs/1909.12285}{{\tt arXiv:1909.12285}}].

\bibitem{Moreno:2019bmu}
E.~A. Moreno, O.~Cerri, J.~M. Duarte, H.~B. Newman, T.~Q. Nguyen, A.~Periwal,
  M.~Pierini, A.~Serikova, M.~Spiropulu, and J.-R. Vlimant, {\it {JEDI-net: a
  jet identification algorithm based on interaction networks}},  {\em Eur.
  Phys. J. C} {\bf 80} (2020), no.~1 58,
  [\href{https://arxiv.org/abs/1908.05318}{{\tt arXiv:1908.05318}}].

\bibitem{Mikuni:2020wpr}
V.~Mikuni and F.~Canelli, {\it {ABCNet: an attention-based method for particle
  tagging}},  {\em The European Physical Journal Plus} {\bf 135} (Jun, 2020).

\bibitem{Cogan:2014oua}
J.~Cogan, M.~Kagan, E.~Strauss, and A.~Schwarztman, {\it {Jet-Images: Computer
  Vision Inspired Techniques for Jet Tagging}},  {\em JHEP} {\bf 02} (2015)
  118, [\href{https://arxiv.org/abs/1407.5675}{{\tt arXiv:1407.5675}}].

\bibitem{deOliveira:2015xxd}
L.~de~Oliveira, M.~Kagan, L.~Mackey, B.~Nachman, and A.~Schwartzman, {\it
  {Jet-images — Deep Learning Edition}},  {\em JHEP} {\bf 07} (2016) 069,
  [\href{https://arxiv.org/abs/1511.05190}{{\tt arXiv:1511.05190}}].

\bibitem{Thaler:2010tr}
J.~Thaler and K.~Van~Tilburg, {\it {Identifying Boosted Objects with
  N-subjettiness}},  {\em JHEP} {\bf 03} (2011) 015,
  [\href{https://arxiv.org/abs/1011.2268}{{\tt arXiv:1011.2268}}].

\bibitem{Thaler:2011gf}
J.~Thaler and K.~Van~Tilburg, {\it {Maximizing Boosted Top Identification by
  Minimizing N-subjettiness}},  {\em JHEP} {\bf 02} (2012) 093,
  [\href{https://arxiv.org/abs/1108.2701}{{\tt arXiv:1108.2701}}].

\bibitem{Gallicchio:2010sw}
J.~Gallicchio and M.~D. Schwartz, {\it {Seeing in Color: Jet Superstructure}},
  {\em Phys. Rev. Lett.} {\bf 105} (2010) 022001,
  [\href{https://arxiv.org/abs/1001.5027}{{\tt arXiv:1001.5027}}].

\bibitem{Faucett:2020vbu}
T.~Faucett, J.~Thaler, and D.~Whiteson, {\it Mapping machine-learned physics
  into a human-readable space},  {\em Physical Review D} {\bf 103} (Feb, 2021).

\bibitem{Montavon_2017}
G.~Montavon, S.~Lapuschkin, A.~Binder, W.~Samek, and K.-R. Müller, {\it
  Explaining nonlinear classification decisions with deep taylor
  decomposition},  {\em Pattern Recognition} {\bf 65} (May, 2017) 211–222.

\bibitem{alber2018innvestigate}
M.~Alber, S.~Lapuschkin, P.~Seegerer, M.~H{{\"a}}gele, K.~T. Sch{{\"u}}tt,
  G.~Montavon, W.~Samek, K.-R. M{{\"u}}ller, S.~D{{\"a}}hne, and P.-J.
  Kindermans, {\it {iNNvestigate Neural Networks!}},  {\em Journal of Machine
  Learning Research} {\bf 20} (2019), no.~93 1--8.
  \url{http://jmlr.org/papers/v20/18-540.html}.

\bibitem{Dasgupta:2013ihk}
M.~Dasgupta, A.~Fregoso, S.~Marzani, and G.~P. Salam, {\it {Towards an
  understanding of jet substructure}},  {\em JHEP} {\bf 09} (2013) 029,
  [\href{https://arxiv.org/abs/1307.0007}{{\tt arXiv:1307.0007}}].

\bibitem{ReLU_1}
X.~Glorot, A.~Bordes, and Y.~Bengio, {\it {Deep Sparse Rectifier Neural
  Networks}},  vol.~15, pp.~315--323, Proceedings of the Fourteenth
  International Conference on Artificial Intelligence and Statistics, JMLR
  Workshop and Conference Proceedings, 2011.
\newblock \url{http://proceedings.mlr.press/v15/glorot11a.html}.

\bibitem{ReLU_2}
V.~Nair and G.~E. Hinton, {\it {Rectified linear units improve restricted
  boltzmann machines}},  p.~807–814, ICML'10: Proceedings of the 27th
  International Conference on International Conference on Machine Learning,
  2010.
\newblock \url{https://dl.acm.org/doi/10.5555/3104322.3104425}.

\bibitem{Sjostrand:2014zea}
T.~Sjöstrand, S.~Ask, J.~R. Christiansen, R.~Corke, N.~Desai, P.~Ilten,
  S.~Mrenna, S.~Prestel, C.~O. Rasmussen, and P.~Z. Skands, {\it {An
  Introduction to PYTHIA 8.2}},  {\em Comput. Phys. Commun.} {\bf 191} (2015)
  159--177, [\href{https://arxiv.org/abs/1410.3012}{{\tt arXiv:1410.3012}}].

\bibitem{Dolen:2016kst}
J.~Dolen, P.~Harris, S.~Marzani, S.~Rappoccio, and N.~Tran, {\it {Thinking
  outside the ROCs: Designing Decorrelated Taggers (DDT) for jet
  substructure}},  {\em JHEP} {\bf 05} (2016) 156,
  [\href{https://arxiv.org/abs/1603.00027}{{\tt arXiv:1603.00027}}].

\bibitem{Shimmin:2017mfk}
C.~Shimmin, P.~Sadowski, P.~Baldi, E.~Weik, D.~Whiteson, E.~Goul, and
  A.~Søgaard, {\it {Decorrelated Jet Substructure Tagging using Adversarial
  Neural Networks}},  {\em Phys. Rev.} {\bf D96} (2017), no.~7 074034,
  [\href{https://arxiv.org/abs/1703.03507}{{\tt arXiv:1703.03507}}].

\bibitem{Kitouni:2020xgb}
O.~Kitouni, B.~Nachman, C.~Weisser, and M.~Williams, {\it {Enhancing searches
  for resonances with machine learning and moment decomposition}},
  \href{https://arxiv.org/abs/2010.09745}{{\tt arXiv:2010.09745}}.

\bibitem{Cacciari:2005hq}
M.~Cacciari and G.~P. Salam, {\it {Dispelling the $N^{3}$ myth for the $k_t$
  jet-finder}},  {\em Phys. Lett.} {\bf B641} (2006) 57--61,
  [\href{https://arxiv.org/abs/hep-ph/0512210}{{\tt hep-ph/0512210}}].

\bibitem{Cacciari:2011ma}
M.~Cacciari, G.~P. Salam, and G.~Soyez, {\it {FastJet User Manual}},  {\em Eur.
  Phys. J.} {\bf C72} (2012) 1896, [\href{https://arxiv.org/abs/1111.6097}{{\tt
  arXiv:1111.6097}}].

\bibitem{Cacciari:2008gp}
M.~Cacciari, G.~P. Salam, and G.~Soyez, {\it The anti-\kt jet clustering
  algorithm},  {\em JHEP} {\bf 04} (2008) 063,
  [\href{https://arxiv.org/abs/0802.1189}{{\tt arXiv:0802.1189}}].

\bibitem{Larkoski:2014wba}
A.~J. Larkoski, S.~Marzani, G.~Soyez, and J.~Thaler, {\it {Soft Drop}},  {\em
  JHEP} {\bf 05} (2014) 146, [\href{https://arxiv.org/abs/1402.2657}{{\tt
  arXiv:1402.2657}}].

\bibitem{PhysRevD.98.030001}
{\bf Particle Data Group} Collaboration, M.~Tanabashi et~al., {\it {Review of
  Particle Physics}},  {\em Phys. Rev. D} {\bf 98} (Aug, 2018) 030001.

\bibitem{adam}
D.~Kingma and J.~Ba, {\it {Adam: A Method for Stochastic Optimization}},  {\em
  International Conference on Learning Representations} (12, 2014)
  [\href{https://arxiv.org/abs/1412.6980}{{\tt arXiv:1412.6980}}].

\bibitem{adam_convergence}
S.~J. Reddi, S.~Kale, and S.~Kumar, {\it {{On the Convergence of Adam and
  Beyond}}},  in {\em International Conference on Learning Representations},
  2018.
\newblock \href{https://arxiv.org/abs/1904.09237}{{\tt arXiv:1904.09237}}.

\bibitem{rene_brun_2019_3895860}
R.~Brun, F.~Rademakers, P.~Canal, A.~Naumann, O.~Couet, L.~Moneta, V.~Vassilev,
  S.~Linev, D.~Piparo, G.~GANIS, B.~Bellenot, E.~Guiraud, G.~Amadio, wverkerke,
  P.~Mato, TimurP, M.~Tadel, wlav, E.~Tejedor, J.~Blomer, A.~Gheata,
  S.~Hageboeck, S.~Roiser, marsupial, S.~Wunsch, O.~Shadura, A.~Bose,
  CristinaCristescu, X.~Valls, and R.~Isemann, {\it root-project/root:
  v6.18/02},  Aug., 2019.

\bibitem{lindsey_gray_2020_4247940}
L.~Gray, N.~Smith, A.~Novak, D.~Taylor, P.~Fackeldey, C.~Carballo,
  P.~Gessinger, J.~Pata, A.~Woodard, Andreas, B.~Fischer, Z.~Surma, A.~Perloff,
  D.~Noonan, L.~Heinrich, N.~Amin, P.~Das, I.~Dutta, J.~Duarte, J.~Rübenach,
  and A.~R. Hall, {\it Coffeateam/coffea: Release v0.6.46},  Nov., 2020.

\bibitem{jim_pivarski_2020_3952728}
J.~Pivarski, P.~Das, C.~Burr, D.~Smirnov, M.~Feickert, T.~Gal, L.~Kreczko,
  N.~Smith, N.~Biederbeck, O.~Shadura, M.~Proffitt, benkrikler, H.~Dembinski,
  H.~Schreiner, J.~Rembser, M.~R., C.~Gu, J.~Rübenach, M.~Peresano, and
  R.~Turra, {\it scikit-hep/uproot: 3.12.0},  July, 2020.

\bibitem{jim_pivarski_2019_3275017}
J.~Pivarski, C.~Escott, M.~Hedges, N.~Smith, C.~Escott, J.~Rembser, J.~Nandi,
  B.~Fischer, H.~Schreiner, P.~Das, P.~Fackeldey, Nollde, and B.~Krikler, {\it
  scikit-hep/awkward-array: 0.12.0rc1},  July, 2019.

\bibitem{tensorflow2015-whitepaper}
M.~Abadi, A.~Agarwal, P.~Barham, E.~Brevdo, Z.~Chen, C.~Citro, G.~S. Corrado,
  A.~Davis, J.~Dean, M.~Devin, S.~Ghemawat, I.~Goodfellow, A.~Harp, G.~Irving,
  M.~Isard, Y.~Jia, R.~Jozefowicz, L.~Kaiser, M.~Kudlur, J.~Levenberg,
  D.~Man\'{e}, R.~Monga, S.~Moore, D.~Murray, C.~Olah, M.~Schuster, J.~Shlens,
  B.~Steiner, I.~Sutskever, K.~Talwar, P.~Tucker, V.~Vanhoucke, V.~Vasudevan,
  F.~Vi\'{e}gas, O.~Vinyals, P.~Warden, M.~Wattenberg, M.~Wicke, Y.~Yu, and
  X.~Zheng, {\it {TensorFlow}: Large-scale machine learning on heterogeneous
  systems},  2015.
\newblock Software available from tensorflow.org.

\end{thebibliography}\endgroup


\end{document}